\documentclass[12pt]{article}  
\usepackage[dvips]{graphicx}
\usepackage{amsmath,amssymb,amsbsy}
\usepackage{amsfonts}
\usepackage{epsfig}
\newcommand{\beq}[1]{ \begin{equation}\label{#1}}
\newcommand{\eeq}{\end{equation}}
\newcommand{\ba}{\begin{array}}

\newcommand{\beqa}{\begin{eqnarray}}
\newcommand{\eeqa}{\end{eqnarray}}
\newcommand{\bear}{\begin{array}{c}}
\newcommand{\bearr}{\begin{array}{cc}}
\newcommand{\ear}{\end{array}}

\newcommand{\mbf}{\mathbf}

\newcommand{\R}{I\kern-.3em{R}}
\newcommand{\no}{\nonumber}
\begin{document}\thispagestyle{empty}
\null
\begin{center}
\vskip 1.cm
{\Large\bf{An analysis of elastic scattering reactions \\
\vskip 0.5cm
with a Fermi-Dirac pomeron opaqueness\\
\vskip 0.5cm
  in impact parameter space}}
\vskip 1.cm
{\bf Claude Bourrely}
\vskip 0.3cm
 Aix-Marseille Universit\'e,
D\'epartement de Physique,\\ Facult\'e des Sciences de Luminy,
13288 Marseille, Cedex 09, France\\
\vskip 0.5cm
{\bf Abstract}
\end{center}
In the Bourrely-Soffer-Wu model (BSW) we introduce for the pomeron
a new opaqueness in impact parameter space in terms of different
quark contributions described by a Fermi-Dirac distribution. 
In order to check the validity of this assumption we consider 
$p~p$, $\bar p~p$, and $\pi^{\pm}~p$ elastic scattering. 
We emphasize the role of the gluon above the diffraction peak in 
the differential cross sections.
Once these contributions are determined we extend the model to
light nuclei elastic reactions like  $p~d$, $p~^4\mbox{He}$ and
$\pi^{\pm}~^4\mbox{He}$.
The results  obtained show a good description of all these elastic
processes over the available experimental energy range 
and  moderate momentum transfer.

\vskip 0.5cm

\vskip 0.5cm
\clearpage
\newpage
\section{Introduction}
\label{intro}
The advent of the LHC collider has renewed the interest of the high energy 
behavior of the $p~p$ elastic scattering and raises the question of the
validity of numerous models devoted to this reaction. Many years ago we
proposed the BSW model \cite{BSW79} and made further developments
\cite{BSW84}-\cite{BSW11} to improve the agreement with experiments.
This model which is based on an impact-picture phenomelogy relies for the
pomeron contribution to the opaqueness
on two assumptions {\it i)} the energy dependence is deduced
from the high-energy behavior of quantum field theory \cite{ttwu1,ttwu2}
 {\it ii)} the momentum transfer dependence follows from the 
supposed proportionality
between the charge density of the proton and the internal distribution
of matter \cite{ttwu3,yang}. With these simple assumptions we were able
to obtain a good description of  the available experimental data obtained
at the ISR, SPS and Tevatron.

To be more precise the assumption  {\it ii)} has led us to take for the
momentum transfer dependence at the Born level a dipole 
in an analogous way to the approximation
made to describe the proton electromagnetic form factor, 
however, this was not sufficient and an extra term was added (see next 
section).
We can stress that the relation between charge and matter density was never
strictly proven, moreover, in this description we ignore the quark 
constituants 
of the proton, so the purpose of the paper is to find a new opaqueness 
expression which involves the proton constituants. 
The key observation is that in BSW
the opaqueness in impact parameter $b$ space is very similar to a Fermi 
function, so we propose that for each quarks we associate 
a Fermi component being dependent on the impact parameter $b$. 
An other justification of this new opaqueness
is provided by our statistical model for
parton distribution functions (PDF) and transverse parton distributions
(TMD) which are built in with Fermi functions \cite{BBS1}, the model
is able to describe a large set of unpolarized and polarized structure 
functions in momentum space.

This idea to introduce a Fermi function  in $b$ space 
has been considered by several authors \cite{islam05}-\cite{galoyan12}
and also in momentum space a Tsallis function 
\cite{cleyman12,cleyman13}.
\footnote{For a review see Ref. \cite{dremin12}.} 
However, most of the authors consider a global opaqueness which 
does not discriminate between the quark components, we will see that in 
our approach the properties of each of them reflect their importance 
inside the proton and that their associated thermodynamical potentials 
remain valid for light nuclei elastic reactions.
Let us also mention the Quark-Diquark model \cite{bialas} and the Generalized
Parton Distributions (GPD) which are function of $b$ and the transverse
momentum $k_T$ \cite{belitsky}.

The paper is organized as follow: after a brief introduction to
the original BSW model in sec. \ref{sumbsw}, we define in sec. \ref{fermopac}
a new expression for the opaqueness. In sec. \ref{ppscat} we analyze the
$p~p$ and $\bar p~p$  elastic scattering data which 
determine the free parameters, the relation between the matter distribution 
and the proton electromagnetic form factor is discussed in sec. \ref{ppemff},
then we extend our approach to $p~d$ in sec. \ref{pdscat} and
to $\mbox{p}~^4\mbox{He}$ elastic scattering in sec. \ref{phescat}.
In order to check the validity of our assumption we consider also
the $\pi~p$ and $\pi~^4\mbox{He}$ elastic scattering
in secs. \ref{pipscat} and  \ref{pihescat} respectively.
The last section contains our conclusion.

\section{A summary of the BSW model}
\label{sumbsw}
In the BSW model \cite{BSW79}-\cite{BSW11} the amplitude is defined  
by the eikonal expression
\beq{ampli}
a(s,t) = \frac{is}{2\pi}\int e^{-i\mbf{q}\cdot\mbf{b}} (1 - 
e^{-\Omega(s,b)})  d\mbf{b} \ ,
\eeq
where the opaqueness
\beq{opac0}
\Omega(s, b ) = S(s)F(b) + R(s,b)\, ,
\eeq
the energy dependence is given by the  complex crossing symmetric
expression deduced from the quantum field theory
\begin{equation}
  \label{eq:Sdef}
  S(s)= \frac{s^c}{(\ln s)^{c'}} + \frac{u^c}{(\ln u)^{c'}}~,
\end{equation}
in Eq. (\ref{opac0}) $F(b)$ is the profile function related to the pomeron 
contribution and $R(s,~b)$ represents  Regge contributions which
are  added to describe the low energy scattering. 
We define at the Born level
the momentum transfer dependence through the product of a dipole multiplied
by an extra function whose property is to avoid spurious dips at large
momentum transfer in the differential cross sections so the 
profile function reads
\beq{formf}
\tilde F(t) = f[G(t)]^2 \frac{a^2 + t}{a^2 -t} \ ,
\eeq
\beq{fgt}
G(t) = \frac{1}{(1 - t/m_1^2)(1 - t/m_2^2)} \ .
\eeq
We will see that this extra function is in fact related to the gluon 
contribution.
The scattering amplitude is expressed as a Bessel transform
\beq{nospin}
a(s,t) = is \int_0^{\infty} J_0(b\sqrt{-t})
(1 - e^{-\Omega(s,\mbf{b})})b  db\,,
\eeq
we notice that the factorization property in Eq. (\ref{opac0}) does not 
hold when the amplitude is eikonalized.
In impact space the profile function reads
\beq{eq:Fsec}
F(b)=\int_0^\infty\,\tilde{F}(t)J_0(b\sqrt{-t}) \sqrt{-t}\,  d\sqrt{-t}\,,
\eeq
where the Bessel transform of $\tilde F(t)$ gives the dimensionless expression
\begin{eqnarray}
&&F(b) = -f m_1^2 m_2^2 \left\{
\left[ 1+ 2a^2 A_{13}\right]A_{12}^2 m_2^2\frac{m_1 b}{2}
K_1(m_1 b)\right. \no \\
&&+\left[ 1+ 2a^2 A_{23}\right]A_{12}^2 m_1^2\frac{m_2 b}{2}
K_1(m_2 b)\no \\
&&+\left[2+ 2 a^2(A_{13}+A_{23})\right]A_{12}^3 m_1^2 m_2^2
\left(K_0(m_1 b) -K_0(m_2 b)\right)\no \\
&&+2 a^2 m_1^2 m_2^2 A_{12} A_{32} A_{31}\times \no \\
&& \times\left[A_{31}\left(K_0(m_1 b) - K_0(a b)\right)
-A_{32}\left(K_0(m_2 b) -K_0(a b)\right)\right] \bigg\}
\label{besstrans}\,,
\end{eqnarray}
the coefficients $A_{ij}$ depend on $m_1,~m_2,~a$.
In Eq. (\ref{besstrans}) the terms associated with the Bessel $K_0(a b)$
which depend on the parameter $a$ introduced in (\ref{formf})
give a negative contribution to the sum.
Conversely
\beq{eq:Fb}
\tilde F(t)=\int_0^\infty F(b) J_0(b\sqrt{-t}) b\, db ~.
\eeq
A fit of experimental data gives for the pomeron parameters the values
\cite{BSW03}
\begin{table}[hpt]
    \centering
    \begin{tabular}{|rllrll|}\hline
$c$ & = & 0.167, & \;\;$c'$&=&0.748\\
$m_1$&=&0.577 GeV,&\;\;$m_2$&=&1.719 GeV\\
$a$&=&1.858 GeV,&\;\;$f$&=&6.971 GeV$^{-2}$\\\hline
    \end{tabular}
    \caption{\label{tab:table1} Parameters of the BSW model.}
\label{table1}
      \end{table}

\section{The Fermi-Dirac opaqueness}
\label{fermopac}
At the level of the BSW Born term in momentum space we used
a modified dipole approximation arguing that there should be some kind
of similarity between the distribution of matter and the
distribution of charge inside the proton.
Taking the Fourier transform of this modified dipole we get the opaqueness 
$\Omega(s,b)$
in the impact parameter space b. Now, looking at the curve 
$F(b)$ in Fig. 2 of Ref. \cite{BSW79}, we observe that its shape can be
approximated by Fermi-Dirac functions.

We know from QCD that inside a proton its constituents are 2 quarks $u$, one $d$, 
a sea and the gluon, we can infer that each of them contribute 
to the profile function $F(b)$ and from our previous observation we deduce 
that their global effect can be described by a Fermi function,
so we make the hypothesis that the individual nature of these constituents is also
of Fermi type and that the sum should reproduce the same profile function $F(b)$
as in BSW.

Now, in analogy with the Fermi PDF expressions which in $Q^2,~x$ space depend on
thermodynamical potentials and a temperature, see Ref. \cite{BBS1}, we propose to
associate to each quark a Fermi function with a thermodynamical potential now 
in $b$ space, namely,
$X_u,~X_d,~X_{\bar q},~X_g$, and a parameter $b_0$ which represents an average size 
localization of the partons inside the proton. For the gluon due to the boson
nature  we use a Bose-Einstein function and  introduce a non zero potential
otherwise its contribution would be infinite for $b = 0$. These properties can be
summarized by the following crossing symmetric expression
\beq{ferm}
F(b) = c_0 \left[\frac{1}{1 + \exp{[\frac{b -X_d}{b_0}]}}
+\frac{c_1}{1 + \exp{[\frac{b -X_u}{b_0}]}} 
 +\frac{c_2}{1 - \exp{[\frac{b +X_g}{b_0}]}} 
+\frac{c_3}{1 + \exp{[\frac{b + X_{\bar q}}{b_0}]}}\right]\,,
\eeq
where the signs in front of the potentials are defined according to the same convention 
as in the case of parton distributions. Here,  $c_0$ plays the role
of the parameter $f$ in BSW,  the coefficients $c_1$, $c_2$ and $c_3$ are the relative 
weight of $u$, $g$ and sea with respect to the quark $d$. We ignore in this first 
approach heavy quarks.

Our goal is to show that the expression
(\ref{ferm}) can be used to describe different elastic reactions and that once
the potentials are determined from $p~p$ elastic scattering their values are  
an {\it intrinsic property of the quarks} also valid for scattering 
reactions involving light nuclei  
and give a reliable description of the proton electric form factor at
low $Q^2$.

\section{The $p~p$ and $\bar p~p$ elastic scattering}
\label{ppscat}
Now, it remains to determine the values of the above parameters by making
a fit of the data. We use the same set of data as in the original 
BSW  \cite{kwak}-\cite{bern86}, 
precisely, the energy ranges from $p_{lab} = 100~\mbox{GeV}$ to 
$\sqrt{s} = 1.8~\mbox{TeV}$ for $p~p$ and
$\bar p~p$, and for the momentum transfer we restrict the values 
to $|t| < 5~\mbox{GeV}^2$.
In order to put more constraints on the pomeron we take into account low energy data
so we use for the Regge contributions the same expressions as in 
BSW \cite{BSW79,BSW03}.
A fit gives for the pomeron parameters the following numerical 
values, Table \ref{table2}:
\begin{table}[hpt]
    \centering
    \begin{tabular}{|rllrll|}\hline
$c$ & = & $0.1677\pm 0.0018$, & \;\;$c'$&=& $0.7103\pm 0.0176$\\
$c_0$ & = & $0.0891\pm 0.0029$, & \;\;$c_1$&=& $13.4678\pm 0.238$\\
$c_2$&=& $21.6197 \pm 0.358$,&\;\;$c_3$&=& $4.9707\pm 0.242$\\
$b_0$&=&$0.3337 \pm 0.0098$ fm,&\;\;$X_u$&=& $0.269\pm 0.0073 $ fm\\
$X_d$&=& $1.0654\pm 0.011 $ fm,&\;\; 
$X_{\bar q}$&=& $1.8837\pm 0.03138$ fm\\
$X_g$&=& $0.6832\pm 0.014$ fm&\;\;&& \\\hline
    \end{tabular}
    \caption{\label{tab:table2} Pomeron parameters of the  Fermi model
for $p~p$ and  $\bar p~p$ elastic scattering.}
\label{table2}
      \end{table}

With these parameters we obtain a $\chi^2 = 2060$ for 955 pts which gives a
$\chi^2/pt = 1.95$ and has to be compared with BSW value $\chi^2/pt \sim 2.8$.
We notice that the parameters $c$, $c'$ are close to the ones obtained
with BSW which means that the asymptotic behavior of $S(s)$ is preserved.

\begin{figure}[ht]
 \vspace*{10mm}
\begin{center}
  \epsfig{figure=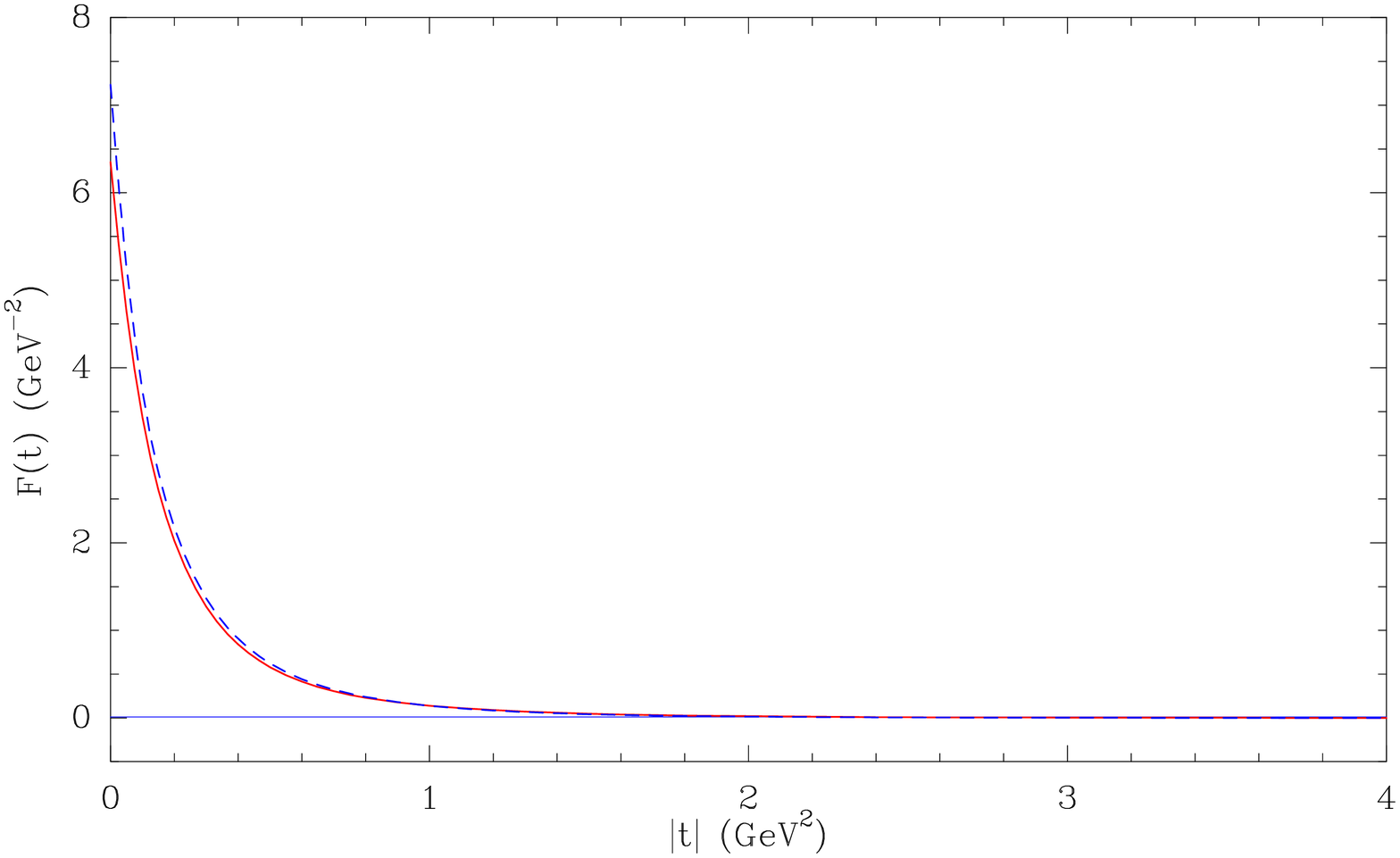,width=10.0cm}
\caption{The profile function $\tilde F(t)$ as a function of $|t|$
for $p-p$ scattering.
Fermi solid red curve, BSW dashed blue curve.}
\label{fig1}
\end{center}
\begin{center}
  \epsfig{figure=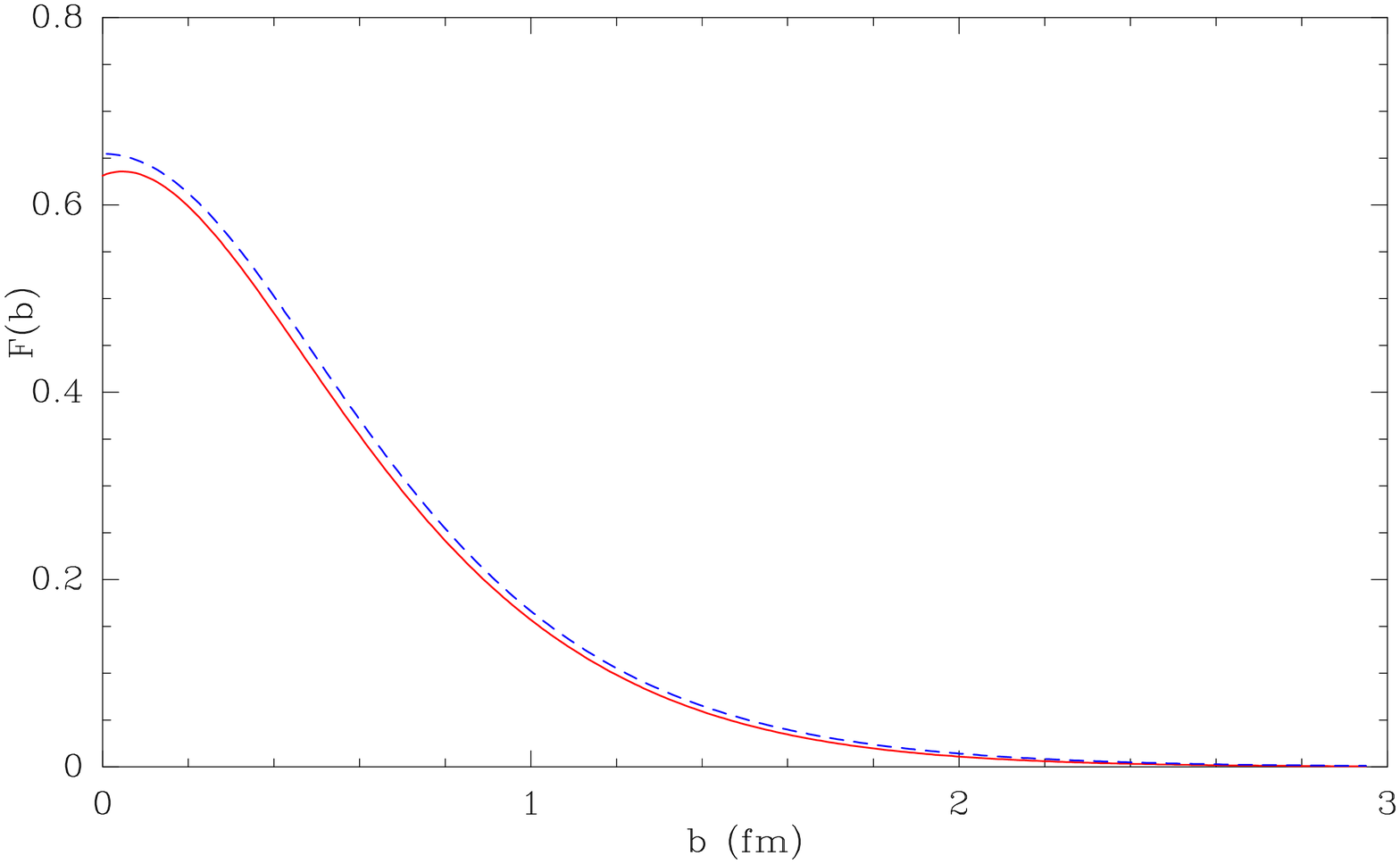,width=10.0cm}
\caption{The profile function $F(b)$ as a function of $b$ for $p-p$ scattering.
 Fermi solid red curve, BSW dashed blue curve.}
\label{fig2}
\vspace*{-1.5ex}
\end{center}
\end{figure}

\begin{figure}[htb]
\vspace*{-18mm}
\begin{center}
  \epsfig{figure=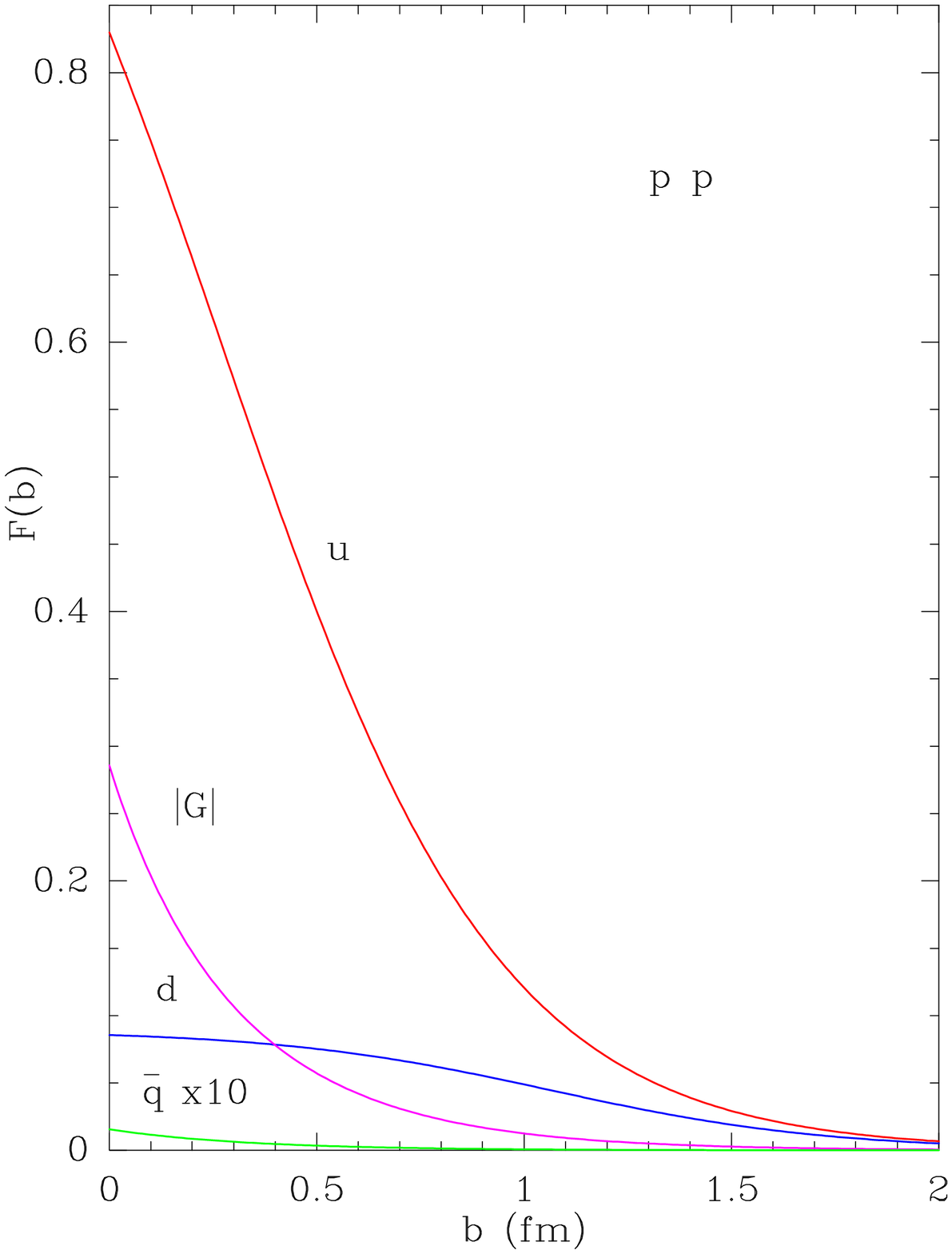,width=7.0cm}
\end{center}
  \vspace*{-5mm}
\caption[*]{Individual contribution of quarks to the profile function
for $p-p$ scattering.}
\label{fi0}
\vspace*{-1.0ex}
\begin{center}
  \epsfig{figure=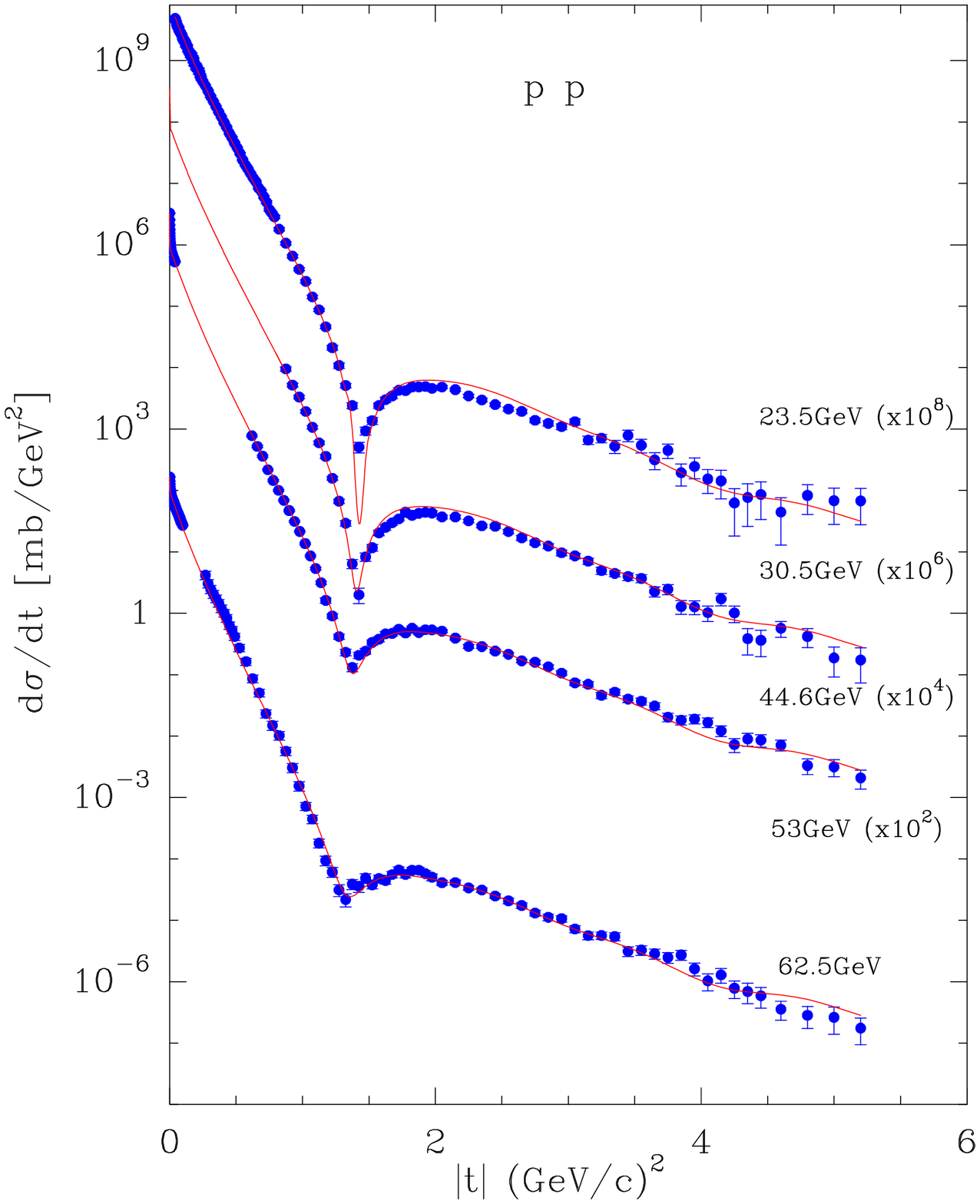,width=8.0cm}
\end{center}
  \vspace*{5mm}
\caption[*]{The $p p$ differential cross section as a function of
$|t|$. 
Experiments from Refs.
\cite{kwak}-\cite{break84}.}
\label{fig4}
\vspace*{-1.0ex}
\end{figure}

We show in Fig.  \ref{fig1} the function $\tilde F(t)$ and in Fig.  \ref{fig2} 
the profile function $F(b)$ produced by the Fermi-Dirac functions both of them
are very close to the BSW curves. With the parameters of Table \ref{table2} 
the individual contribution of quarks to the profile function is shown 
in Fig. \ref{fi0}, we see that they
are concentrated around 1 fermi which is the expected size of the proton,
the two quarks $u$ in the proton give the main contribution compared to the
$d$ quark, the gluon has a contribution concentrated at small $b$.

\begin{figure}[hpb]
\vspace*{-18mm}
\begin{center}
  \epsfig{figure=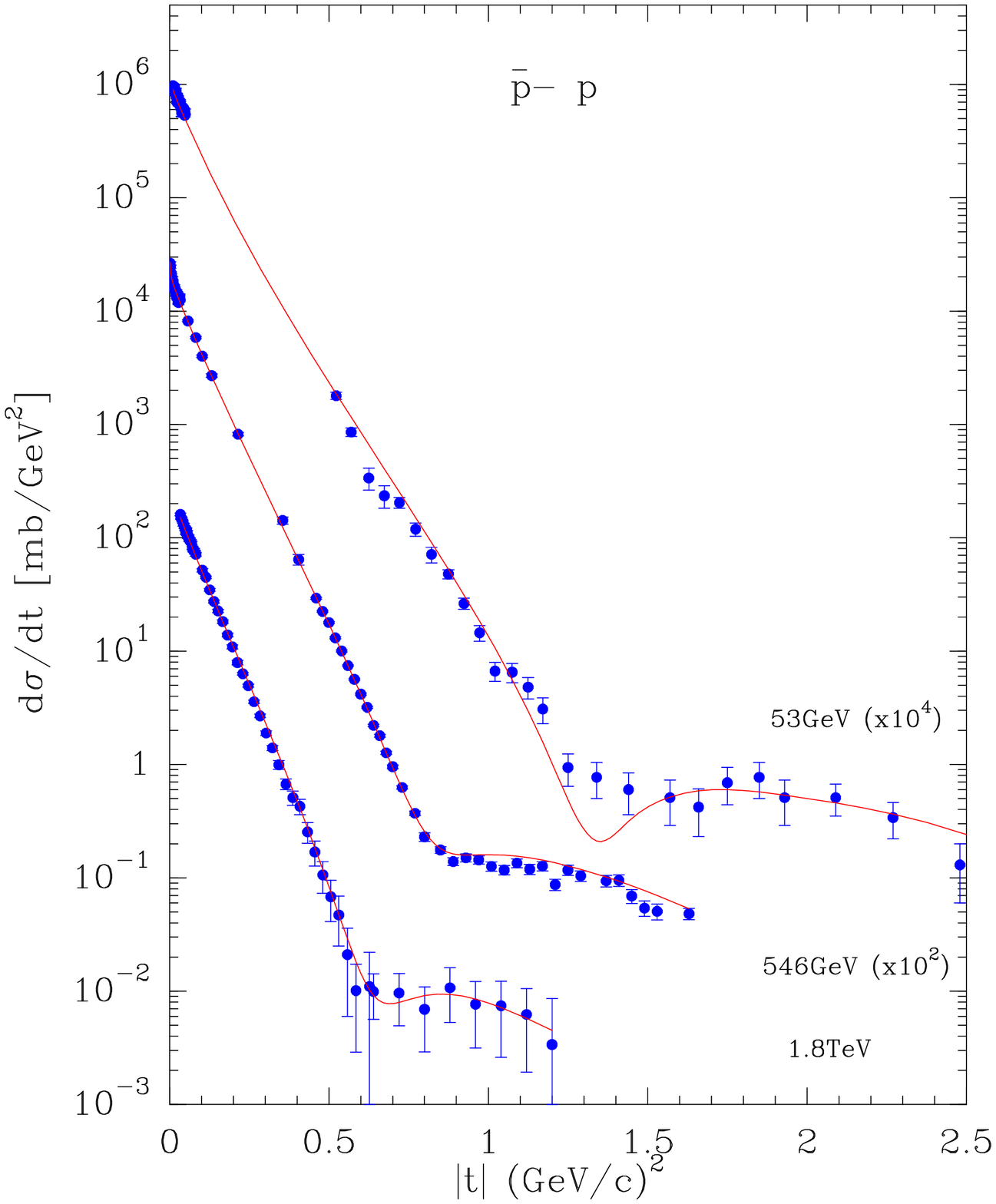,width=8.0cm}
\end{center}
  \vspace*{-5mm}
\caption[*]{The $\bar p p$ differential cross section as a function of
$|t|$.
Experiments from Refs.
\cite{break85,ambro82},  \cite{aug93}-\cite{bern86}.}
\label{fig5}
\begin{center}
  \epsfig{figure=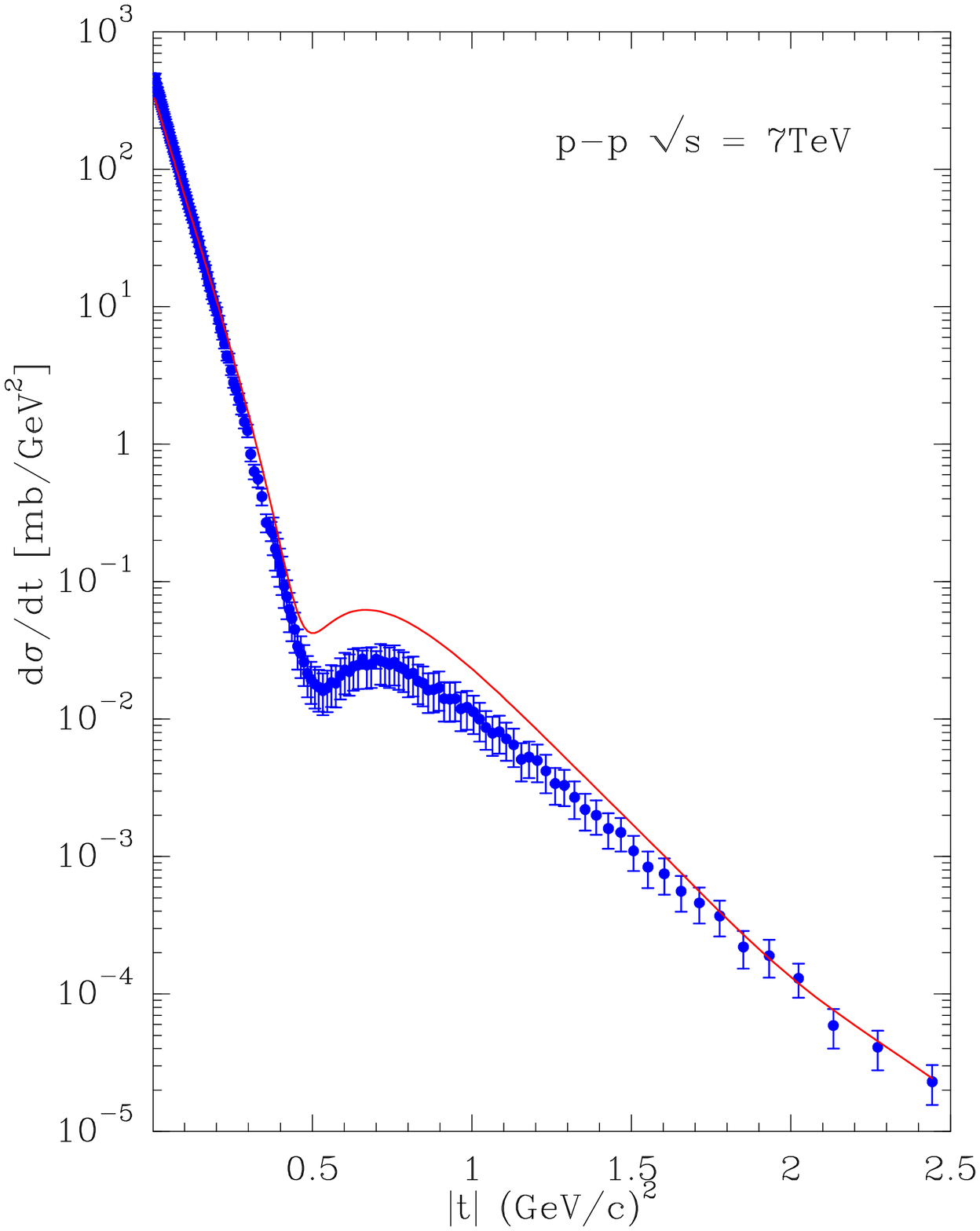,width=9.0cm}
\end{center}
  \vspace*{-5mm}
\caption[*]{A prediction of the Fermi model compared to the TOTEM experimental
data \cite{totem}.}
\label{fi3}
\vspace*{-1.0ex}
\end{figure}

\begin{figure}[hbp]
\begin{center}
  \epsfig{figure=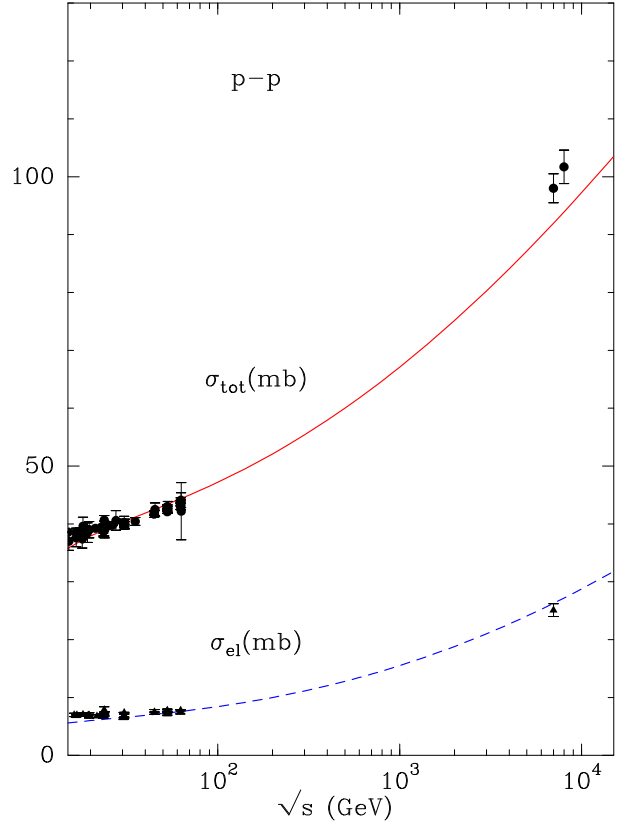,width=8.0cm}
\caption{The p-p total and elastic cros sections as a function
of $\sqrt{s}$. Experimental data from Refs. \cite{totem,pdg}.}
\label{fig7}
\end{center}
\end{figure}

We plot in Figs \ref{fig4}-\ref{fig5} the differential cross sections for $p~p$ 
and 
$\bar p~p$ where we obtain  a good agreement with the data.
A prediction at  $\sqrt{s} = 7$ TeV shows that the Fermi version presents as 
BSW  the same mismatch at large $t$ when compared with the TOTEM
differential cross section measurement \cite{totem}, see Fig. \ref{fi3}. 
At this energy we predict $\sigma_{tot} = 91.95 \pm 1.2$ mb, 
$\sigma_{el} = 25.7\pm 2.2$ mb, $\rho$ = 0.124,
these values have to be compared with the TOTEM data  
$\sigma_{tot} =98.0\pm 2.5$ mb, notice that we agree with 
$\sigma_{el} = 25.43\pm 1.07$ mb and the position of the first minimum  
of the differential cross section, we show in Fig. \ref{fig7} the behavior 
of the total and elastic cross sections.

Let us make a comment: the inclusion the TOTEM data in our
fit notably increase the $\chi^2$, so the quoted parameters values are
obtained leaving aside these data.
A discussion of the BSW model with respect to the TOTEM data is reported 
in Ref. \cite{kohara12}.
In fact, our pomeron whose energy dependence is controlled by the
parameters $c, c'$ which  are constrained by a fit in an energy range 
from low energy up to 1.8 TeV cannot give a total cross section 
as high as the one obtain by TOTEM. In order to reach this value
we need a revision of the pomeron behavior, but before doing any modification
we wait for a confirmation from an other experiment \cite{ALFA}.
Let us point out that at the Tevatron energy 1.8 TeV we obtain
$\sigma_{tot} = 73.6 \pm 1.5$ mb which is in agreement within the experimental
range $71.42 \le \sigma_{tot} \le 80.03$ mb with an average error 2.4 mb
\cite{amos92}-\cite{avila02}.
Notice that the values of the above parameters   $c, c'$ are perfectly 
compatible with the high energy behavior of light nuclei reactions discussed 
in the next sections.

\noindent {\bf{ The role of the gluon}}\\
In the Born term Eq. (\ref{formf}) of BSW we have introduced the extra term
$\frac{a^2 + t}{a^2 -t}$ in order to cancel a spurious second dip
in the differential cross section, 
this term implies that $\tilde F(t)$ has a zero at 
$|t| = a^2 = 3.74~\mbox{GeV}^2$ and becomes negative above. 
The Bessel transform of the Fermi distribution (\ref{ferm}) 
with respect to $b$ gives a function $\tilde F(t)$ which 
has also a zero at $|t| =  4.3~\mbox{GeV}^2$ and a negative value above, 
see Fig. \ref{fig1}. We will show that the origin of this zero 
is produced in fact by the gluon as we now explain.

Looking at the expression of $F(b)$ Eq. (\ref{ferm}) it contains 4 terms 
including the $u$, $d$, the sea and the gluon contributions,
let us suppose that we remove the gluon contribution, a fit made with only the
$u$ and $d$ and the sea gives a very large $\chi^2$, moreover, $\tilde F(t)$ 
has no zero, so one can conclude that the gluon contribution is 
necessary to obtain a reasonable $\chi^2$ and to produce 
 a zero in   $\tilde F(t)$.
Concerning the gluon a more detailed comparison between the
Fermi and the BSW approaches can be made. When making a plot of
Eq. (\ref{besstrans}) we observe that the
terms associated with Bessel functions whose arguments depends on $m_1$
or $m_2$ give a positive contribution to $F(b)$, while
terms associated with the parameter $a$ give a negative contribution.
In the Fermi case the gluon has a denominator $1 - \exp{[\frac{b +X_g}{b_0}]}$
where the minus sign reflects the Bose nature of the contribution,
now the value of the potential $X_g$ must be such 
that the denominator never vanishes otherwise we get a singularity,
taking also into account the constraint for $b = 0$ we see that the denominator
must be always negative which makes a clear correspondence
between the gluon and the
contribution due to the term associated with the parameter $a$ in BSW.
\clearpage
\newpage
\clearpage
\newpage

\section{ The proton  electric form factor at low $Q^2$}
\label{ppemff}
In section \ref{sumbsw}, we have introduced the BSW profile function 
$\tilde F(t)$ Eq. (\ref{formf}) which depends on $G^2(t)$ (Eq. \ref{fgt})
interpreted as a nuclear form factor. In section \ref{fermopac},
we have defined a new $\tilde F(t)$ as the Bessel transform of the Fermi-Dirac 
expressions (\ref{ferm}).
Now, we raise the question if 
there is any relation between this nuclear form factor
and the electromagnetic form factor of the proton. To this end, we define
using Eq. (\ref{ferm})
the proton electric form factor  by the expression
\beqa
G^2_e(Q^2) &=& \int_0^{\infty} bdb J_0(bQ)  f^2_{e}
\left[\frac{1}{1 + \exp{[\frac{b -X_d}{b_e}]}}
+\frac{c_1}{1 + \exp{[\frac{b -X_u}{b_e}]}} \right.\no \\
& & \left. +\frac{c_2}{1 - \exp{[\frac{b +X_g}{b_e}]}} 
+\frac{c_3}{1 + \exp{[\frac{b + X_{\bar q}}{b_e}]}}\right]\,.
\label{formfac}
\eeqa
Compared to Eq. (\ref{ferm}) we introduce the normalization factor
$f^2_e$ and replace the  quarks extension $b_0$ inside  the proton  by
$b_e$ which corresponds to the electromagnetic case, all the other parameters
are kept fixed at the values given in Table \ref{table2}.
 The normalisation factor
$f^2_e$ is determined by the condition $G_e(0) = 1$, we obtain
$f^2_e = 0.0143~\mbox{GeV}^{2}$ and  the best agreement with the
experimental form factor data gives
$b_e = 0.326$ fm a value slightly less
than $b_0 = 0.337$ fm, we can interpret this small  difference from the 
fact that u quarks  give the most important contribution at small $b$ 
(see Fig. \ref{fi0}) and carry 4/3 of the charge while the d quark giving a
smaller contribution has a charge -1/3.

\begin{figure}[htp]
\begin{center}
  \epsfig{figure=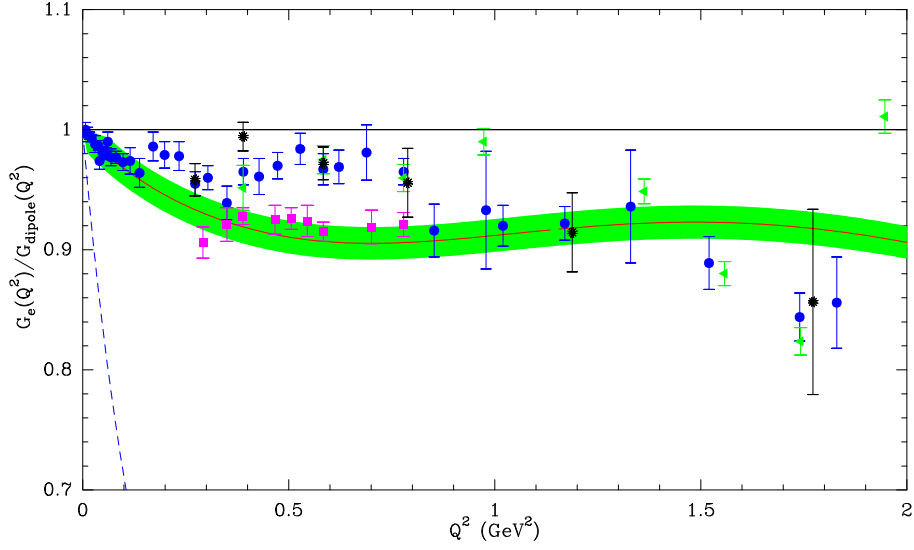,width=12.0cm}
\caption{The proton electric form factor $G_e(Q^2)$
normalized to $G_{dipole}$ as a function of $Q^2$.
Fermi solid curve, uncertainty domain  shaded area, BSW dashed curve.
Experimental data: square   \cite{zhan10}, triangle \cite{berger71},
star \cite{hanson73}, circle a data analysis presented in Ref. \cite{arring07}.}
\label{figformfac}
\end{center}
\end{figure}

In order to make a comparison with experiment we have to rely
on the available measured ratio $G_e(Q^2)/G_{dipole}(Q^2)$\footnote{
$G_{dipole}(Q^2) = \frac{1}{(1 +Q^2/0.71)^2}$ is the usual dipole form factor.}
since we are not able to compute the magnetic form factor,so  we cannot use the
measurements $G_e/G_m$.
In Fig. \ref{figformfac} we show the plot $G_e(Q^2)/G_{dipole}(Q^2)$ produced
by the Fermi distributions (\ref{formfac}) (solid line), the agreement with 
experimental data at low $Q²$ is relatively good, 
we notice that the recent polarized experiment at JLab 
\cite{zhan10} (squares in the figure) gives the most precise values. 
For reference we show the case of BSW given by Eq. (\ref{formf})
we observe a fast decrease of this ratio because the parameters $m_1$ and
$m_2$ are only valid in the nuclear case.

In the introduction we raised the question of a possible relation between the
nuclear and electromagnetic form factors, our Fermi approach shows clearly that
with only two new parameters we can make a close link between the distribution of 
matter and the distribution of charge inside the proton.
\section{The $p~d$ elastic scattering}
\label{pdscat}
In the previous section we  considered the elastic scattering
between two elementary particles $p~p$, $\bar p p$ and found the
basic properties of quarks, sea and gluon  interaction through a Fermi-Dirac 
function in impact parameter space. The question arises how to extend this 
type of interaction when a light nucleus like the deuteron is involved
in the $p~d$ elastic scattering.

Our theoretical input for the profile function is the same formula 
defined for $p~p$ by Eq. (\ref{ferm}) where we keep the 
same value of the parameters $c, c'$ and the thermodynamical potentials, 
the only free parameters are the normalization coefficients 
$c_0, c_1, c_2, c_3$, and the parameter $b_0$ associated with 
the deuteron size, we infer that since the total cross section for this 
process is higher than in the proton case $c_0$ must increase.

\begin{figure}[htb]
\vspace*{8mm}
\begin{center}
  \epsfig{figure=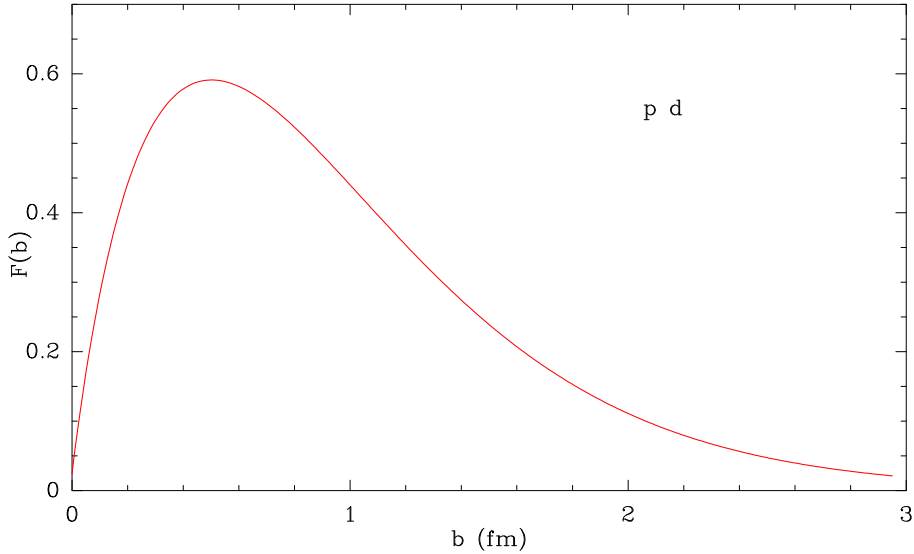,width=12.0cm}
\caption{The profile function $F(b)$ as a function of $b$.}
\label{figbpdeut}
\end{center}
\end{figure}

\begin{figure}[hp]
\begin{center}
  \epsfig{figure=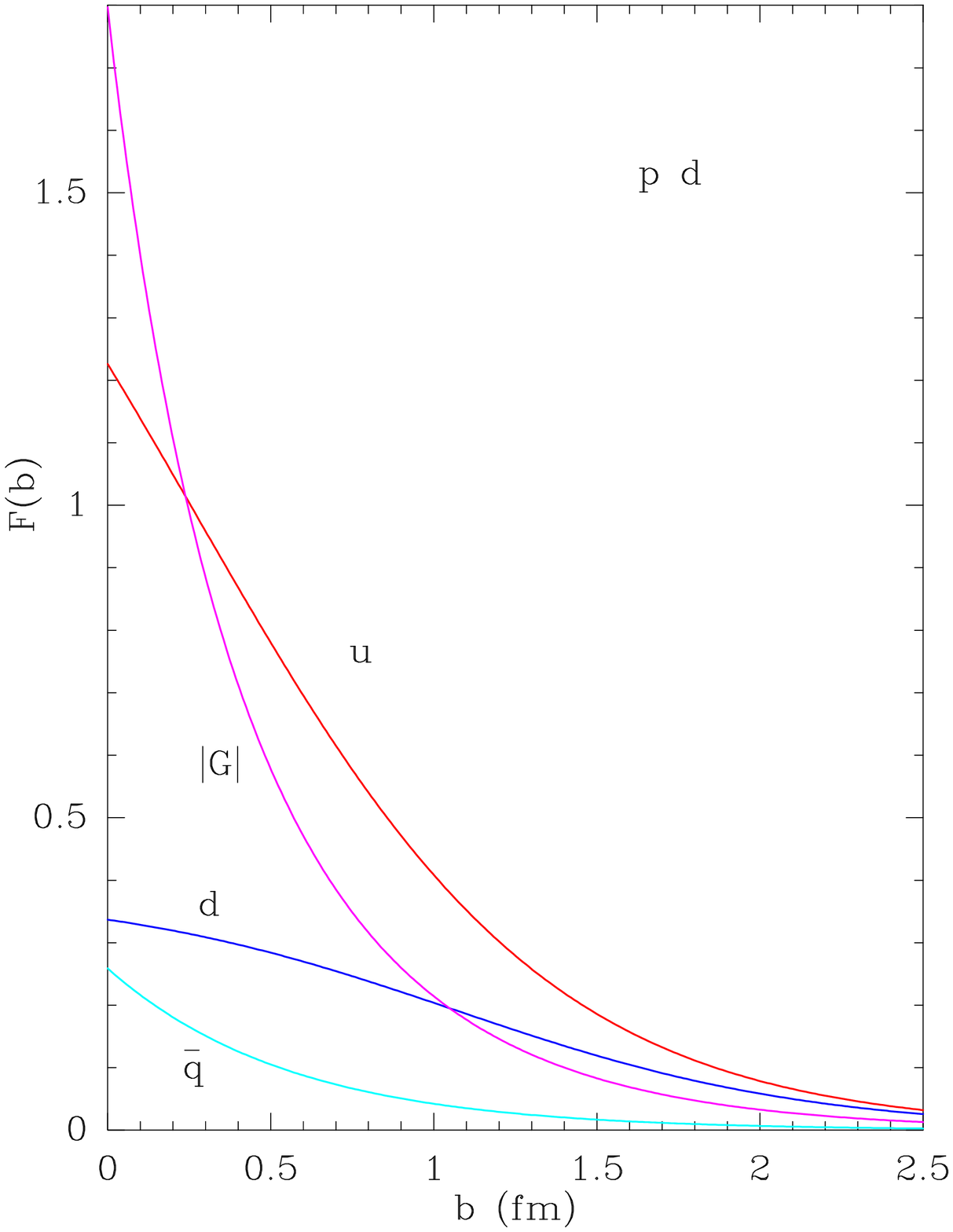,width=8.0cm}
\end{center}
  \vspace*{-5mm}
\caption[*]{Individual contribution of quarks to the profile function
for $p-d$.}
\label{fi0pdeut}
\end{figure}

\begin{figure}[htp]
\vspace*{-18mm}
\begin{center}
  \epsfig{figure=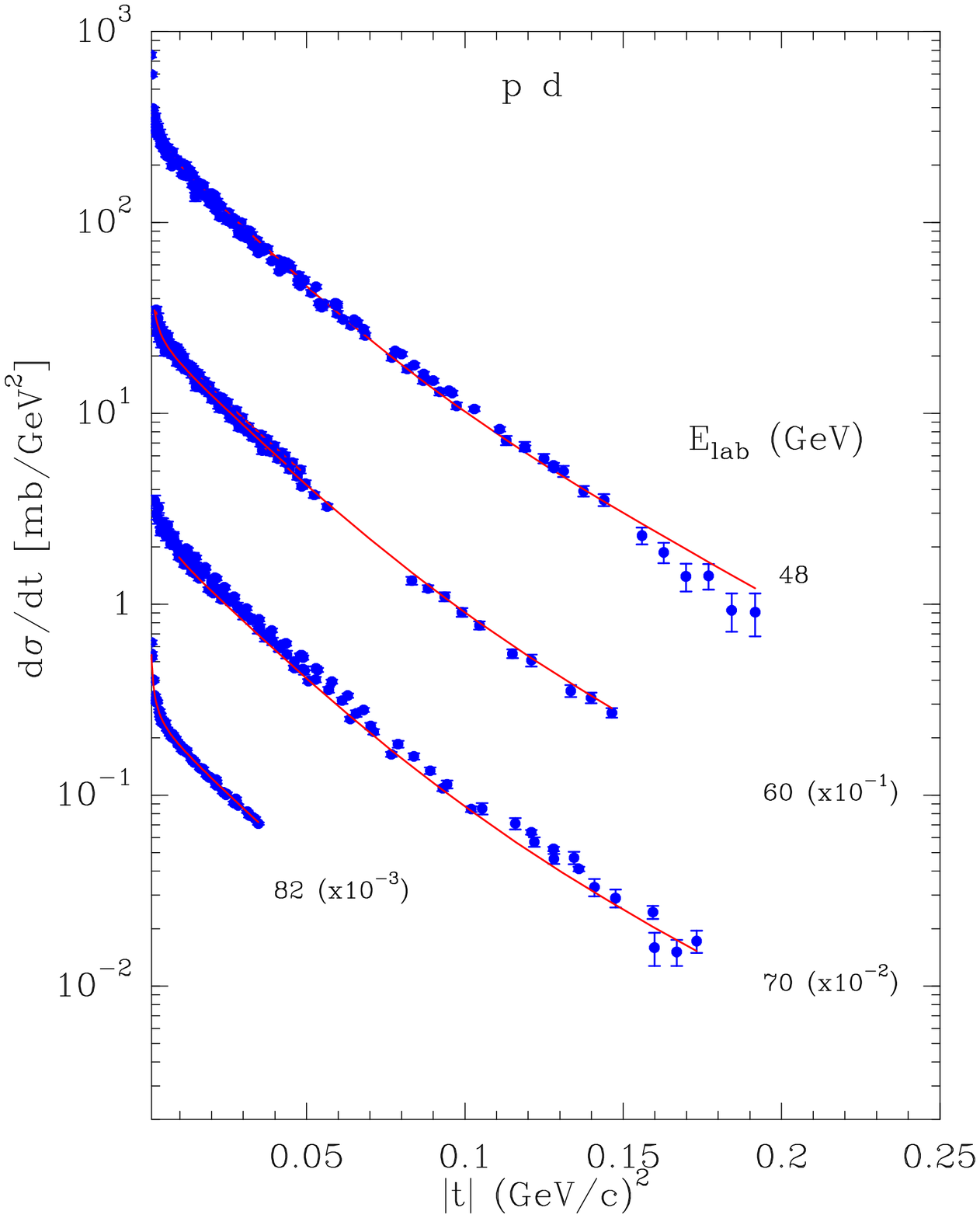,width=8.0cm}
\end{center}
  \vspace*{-5mm}
\caption[*]{The $p d$ differential cross section as a function of $|t|$.
Experiments from Refs.
\cite{akimov75}-\cite{gross78}.}
\label{fi1pdeut}
\vspace*{-1.0ex}
\begin{center}
  \epsfig{figure=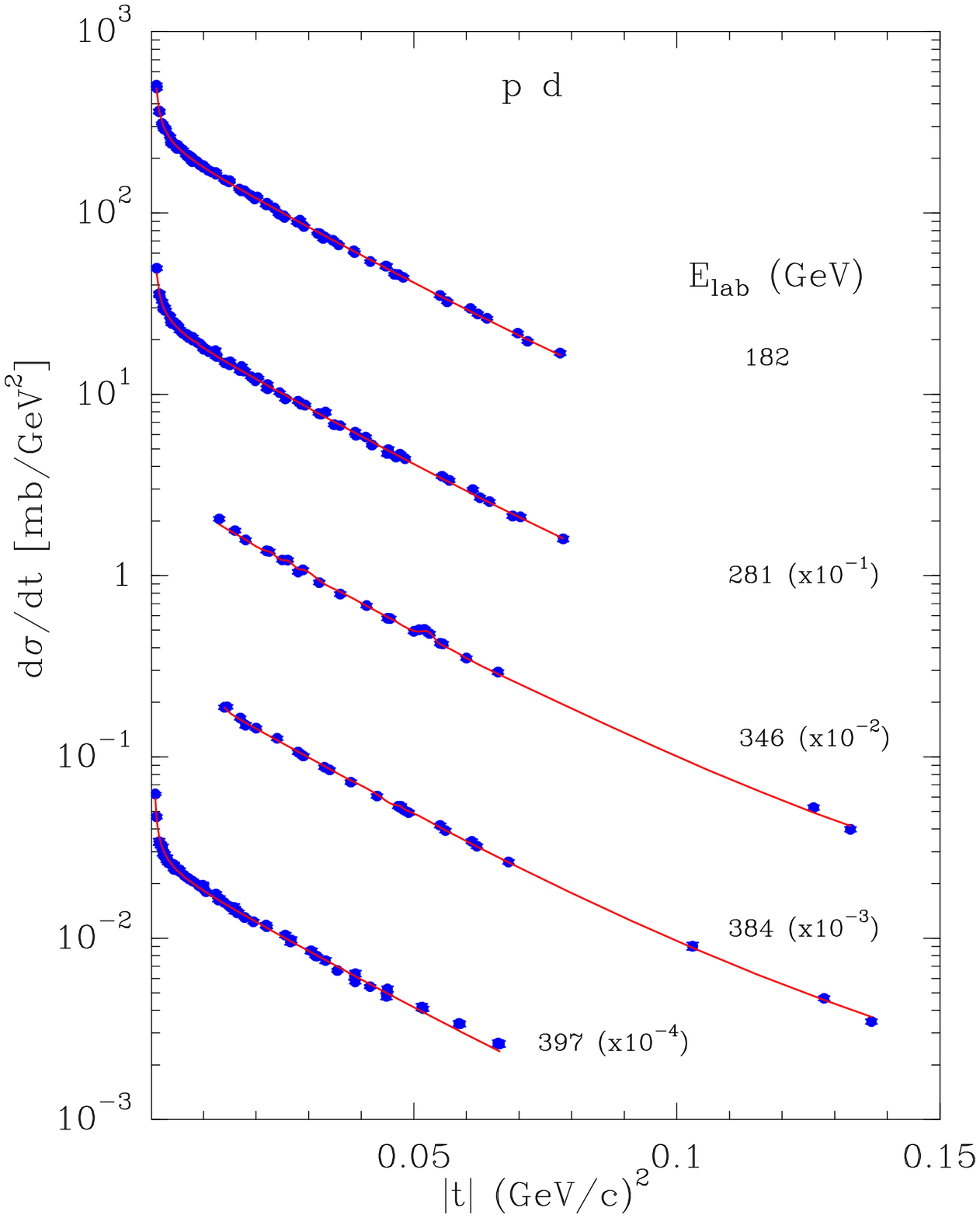,width=8.0cm}
\end{center}
  \vspace*{-5mm}
\caption[*]{The $p d$ differential cross section as a function of $|t|$ 
continued.}
\label{fi2pdeut}
\vspace*{-1.0ex}
\end{figure}

The experimental data \cite{akimov75}-\cite{gross78} cover the energy range 
$40 \le p_{lab} \le 397$ GeV 
and the momentum transfer range $0.00077 \le |t| \le 0.2435$ GeV$^2$
\cite{akimov75}-\cite{gross78}. 
Of course, the energy domain is more limited than in the $p~p$ reaction, 
and  the momentum transfer covers only low $|t|$ values, nevertheless,
we find interesting to check the validity of our assumption on the 
universality of the thermodynamical potentials in this case.
After a fit of data
we obtain  a $\chi^2 = 1533$ for 1000 pts giving a $\chi^2/pt = 1.53$
which is  slightly better than the proton value.
The resulting parameters for the pomeron are given in Table \ref{tab:table6}.
\clearpage
\newpage
\begin{table}[htbp]
    \centering
    \begin{tabular}{|rllrll|}\hline
$c_0$ & = & $0.0726\pm 0.002$, & \;\;$c_1$&=& $33.1219\pm 0.229$\\
$c_2$&=& $7.4228\pm 0.0.23$,&\;\;$c_3$&=& $-35.592\pm 1.09$\\
$b_0$&=& $0.544\pm 0.0122$ fm,&\;\;&& \\ \hline
    \end{tabular}
    \caption{\label{tab:table6} Pomeron parameters of the Fermi model
for $p~d$  elastic scattering.}
      \end{table}
The profile function $F(b)$  shown in Fig. \ref{figbpdeut} differs
 from the $p~p$ case with a maximum at  $b = 0.5$ fm.
In Fig. \ref{fi0pdeut} we plot the different
components of the pomeron contribution, we observe that the gluon and the
quark u give the major contribution. The differential
cross sections show a perfect agreement with the data in the measured
low t region, see Figs. \ref{fi1pdeut}-\ref{fi2pdeut}.
Concerning the total cross we obtain for instance at $E_{lab} = 240~\mbox{GeV}$,
$\sigma_{tot} = 73.77\pm 0.4$ mb to be compared with the experimental value
$74.42 \pm 0.53$ mb \cite{caroll79}

This result confirms that our basic Fermi interaction between quarks obtained
in the elastic proton case where we have a system made of $4u + 2d$
remains valid for this light nucleus scattering where now it  contains 
$5u +4d$. 

\clearpage
\newpage
\section{The $\mbox{p}~^4\mbox{He}$ elastic scattering}
\label{phescat}
Following the same approach of the previous sections,
we propose to describe the elastic reaction $\mbox{p}~^4\mbox{He}$ from
the measurements made at Fermilab with a gas target in a range
of energies from 97 to 400 GeV and momentum transfer 
$0.003 \leq |t| \leq 0.52~\mbox{GeV}^2$ \cite{bujak81,burq81}. 

Our theoretical input for the profile function relies on the same formula defined 
for $p~p$ by Eq. (\ref{ferm}) where we keep the same value 
of the parameters $c, c'$ and the thermodynamical potentials, 
here again the only free parameters are the 
 coefficients $c_0, c_1, c_2, c_3$  and the parameter
$b_0$.

A fit gives  a $\chi^2 = 476$ for 504 pts or a $\chi^2/pt = 0.94$.
The resulting parameters for the pomeron are given in Table \ref{tab:table4}:
\begin{table}[hp]
    \centering
    \begin{tabular}{|rllrll|}\hline
$c_0$ & = & $0.0134\pm 0.0015$, & \;\;$c_1$&=& $29.9041\pm 1.01$\\
$c_2$&=& $24.9568\pm 1.02$,&\;\;$c_3$&=& $0.4968\pm 0.067$\\
$b_0$&=& $0.5967\pm 0.0018$ fm &\;\;&& \\ \hline
    \end{tabular}
    \caption{\label{tab:table4} Pomeron parameters of the Fermi model
for $\mbox{p}~^4\mbox{He}$  elastic scattering.}
      \end{table}

\begin{figure}[htb]
\vspace*{-18mm}
\begin{center}
  \epsfig{figure=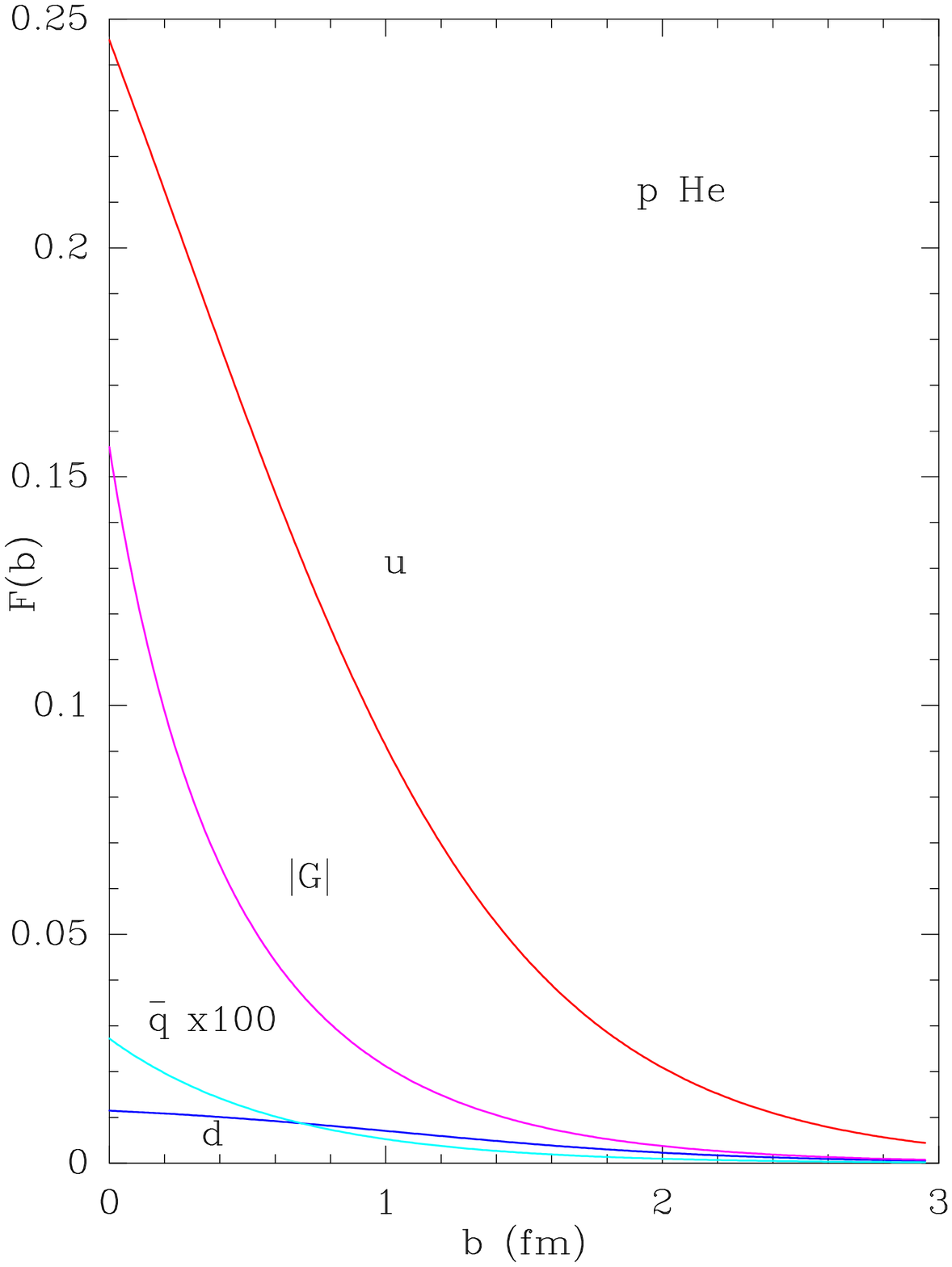,width=8.0cm}
\end{center}
  \vspace*{-5mm}
\caption[*]{Individual contribution of quarks to the profile function
for $\mbox{p}~^4\mbox{He}$ .}
\label{fiphe0}
\vspace*{-0.5ex}
\begin{center}
  \epsfig{figure=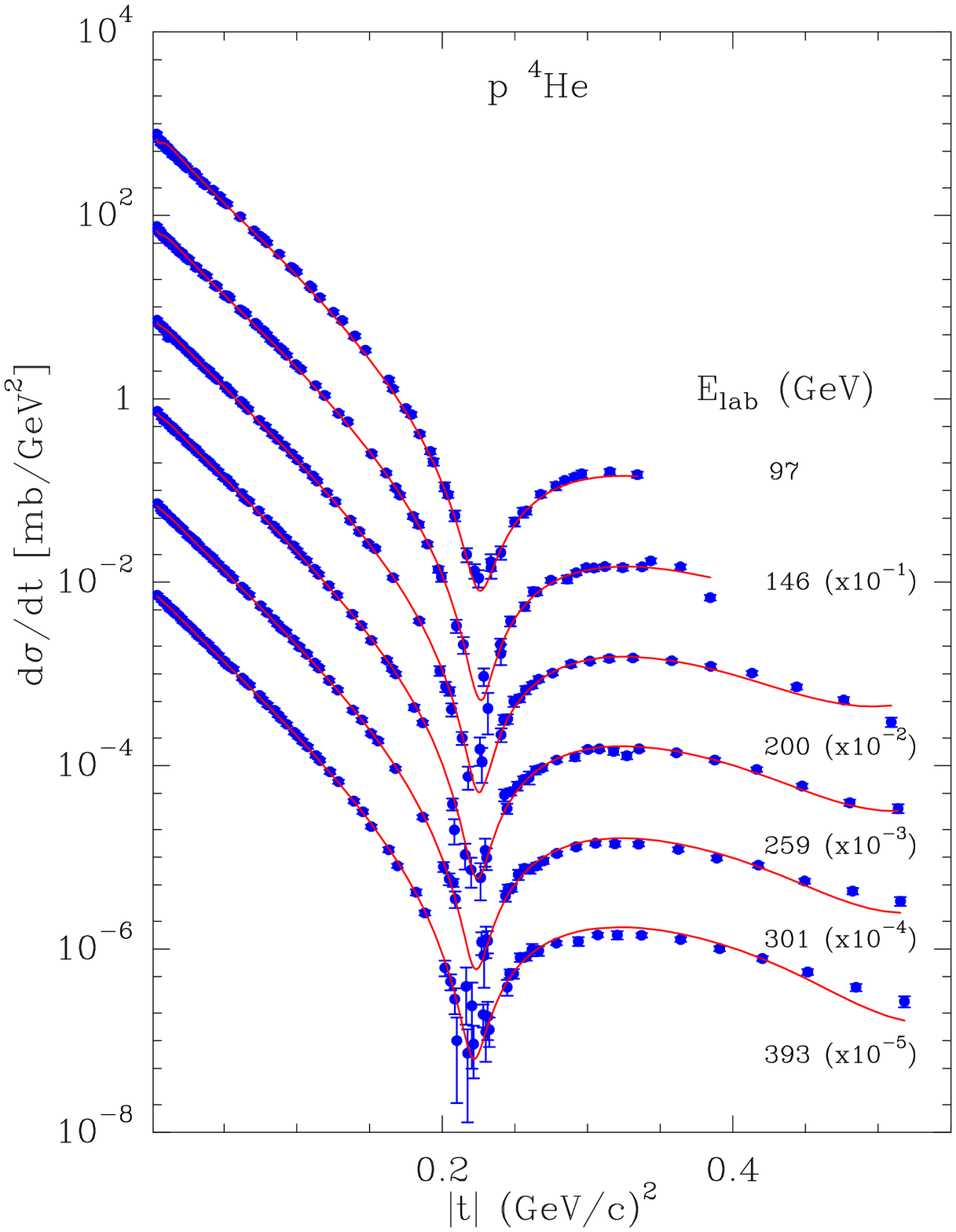,width=8.0cm}
\end{center}
  \vspace*{-5mm}
\caption[*]{The $\mbox{p}~^4\mbox{He}$ differential cross section as
a function of $|t|$.
Experiments from Refs.  \cite{bujak81,burq81}.}
\label{fiphe1}
\vspace*{-1.0ex}
\end{figure}

With these parameters we plot in Fig. \ref{fiphe0} the different
components of the pomeron contribution, we observe that the gluon and the
quark $u$ give the major contribution
a situation similar to the $p~d$ case (see  Fig. \ref{fi0pdeut}).
A plot of the  differential cross sections is shown in  Fig. \ref{fiphe1}, 
we notice that the dip region
is well described, concerning the total cross we obtain at $E_{lab} = 250$ GeV
$\sigma_{tot} = 132.13 \pm 0.5$ mb to be compared with 
$\sigma_{tot} = 131.6 \pm 0.8$ mb of Ref. \cite{burq81}.
For this reaction the agreement with the data validates our assumption
on the structure of the profile function $F(b)$ and the fact that the
thermodynamical potentials are kept the same. 

\clearpage
\newpage

\section{The $\pi^{\pm}~p$ elastic scattering}
\label{pipscat}
In addition to $p~p$ scattering the $\pi^{\pm}~p$ must give new informations
on the partons content of the $\pi~p$ interaction.
For this reaction, in the original BSW  the pomeron contribution is defined by the 
expression:
\beq{formfpi}
\tilde F(t) = f_{\pi} G(t) \mathcal F_{\pi}(t)\frac{a_{\pi}^2 + t}{a_{\pi}^2 -t}\ ,
\eeq
where $\mathcal F_{\pi}(t) =\frac{1}{1-t/m_{3\pi}^2}$ is a simple pole, 
from a fit we obtained the following parameters
\beq{parampi}
m_{3\pi} = 0.7665~\mbox{GeV}, \quad  f_{\pi} =4.2414, \quad  a_{\pi} = 2.3272 
~\mbox{GeV} \,.
\eeq
For the Fermi description of the $\pi~p$ interacting system in impact parameter 
space we explore a slightly different approach compared to the proton case, 
in the sense that we introduce a different quark potential according 
to the charge of the pion so that the new pomeron profile function takes the form
\beq{fermpi}
F^{\pm}_{\pi}(b) = d_0 \left[\frac{1}{1 + \exp{[\frac{b -X^{\pm}_d}{b_0}]}}
+\frac{d_1}{1 + \exp{[\frac{b -X^{\pm}_u}{b_0}]}} 
 +\frac{d_2}{1 - \exp{[\frac{b +Z_g}{b_0}]}} 
+\frac{d_3}{1 + \exp{[\frac{b + Z_{\bar q}}{b_0}]}}\right]\,.
\eeq
Since the quark structure for $\pi^+$ is $u~\bar d$ and for $\pi^-$ is 
$d~\bar u$, we define a set of thermodynamical potentials
$X^{\pm}_u,~X^{\pm}_d$ corresponding to $\pi^{\pm}$, the reason being that
in the system at rest we have $3u + d$ for  $\pi^+$  and $2u +2d$ for
$\pi^-$ so the potentials are not necessarily to be the same as in p-p.
For the sea we introduce  a global potential $Z_{\bar q}$, 
and for the gluon component a potential $Z_g$.
The  parameters $c,~c'$ which drive the asymptotic energy  behavior
are kept  the same as in  $p~p$ (see Table \ref{tab:table2}).

A simultaneous fit of $\pi^{\pm}$ data for 
$p_{lab} = 100-250$ GeV, and momentum transfer  $|t| < 2.5~\mbox{GeV}^2$
 \cite{aker76}-\cite{adam87},
gives a $\chi^2 = 1005$ with 608 pts or a $\chi^2/pt = 1.65$.
The resulting parameters for the pomeron are given in Table \ref{tab:table3}.
\begin{table}[hp]
    \centering
    \begin{tabular}{|rllrll|}\hline
$d_0$ & = & $3.4408\pm 0.118$, &\;\; $d_1$&=& $2.406\pm 0.21$ \\
$d_2$&=& $2.5345\pm 0.195$,&\;\; $d_3$&=& $5.5128\pm 0.323$  \\
$X^{+}_{u}$&=& $0.2802\pm 0.002$ fm,&\;\;$X^{-}_{u}$&=& $0.2307\pm 0.0197$ fm   \\
$X^{+}_{d}$&=& $0.0096\pm 0.0004$   fm, &\;\;$X^{-}_{d}$&=&$ 0.1772\pm 0.0065$ fm \\
$Z_{\bar q}$&=& $0.6323\pm 0.0118$ fm,&\;\; $Z_g$&=& $0.3537\pm 0.0116$ fm \\
$b_0$&=& $0.3096\pm 0.004$ fm&\;\;&&\\\hline
    \end{tabular}
    \caption{\label{tab:table3} Pomeron parameters of the Fermi model
for $\pi^{\pm}~p$ elastic scattering.}
      \end{table}

\begin{figure}[htbp]
\begin{center}
  \epsfig{figure=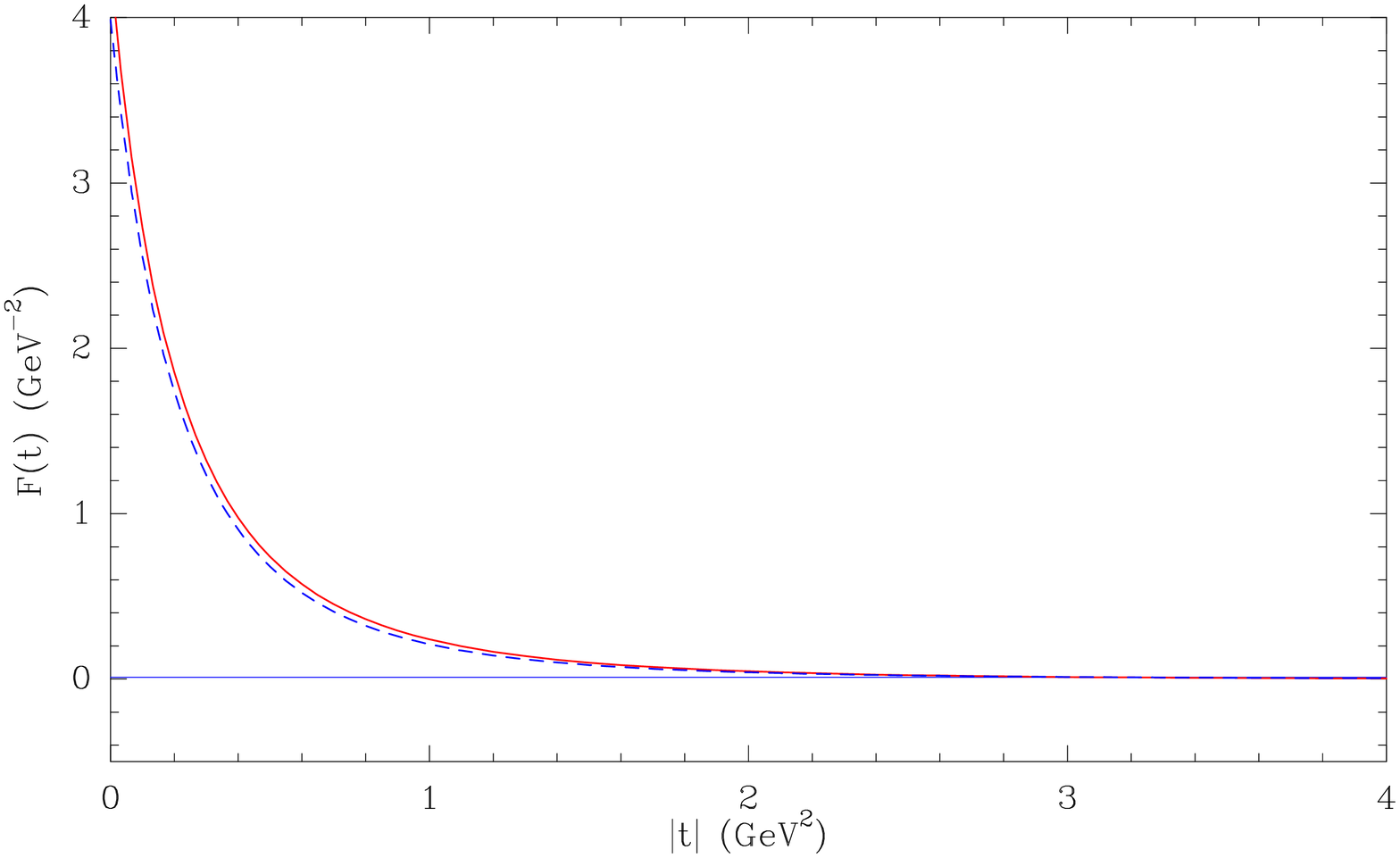,width=12.0cm}
\caption{The profile function $\tilde F(t)$ as a function of $|t|$
for $\pi^{\pm}~p$ .
 Fermi solid red curve, BSW dashed blue curve.}
\label{fig1pi}
\end{center}
 \vspace*{8mm}
\begin{center}
  \epsfig{figure=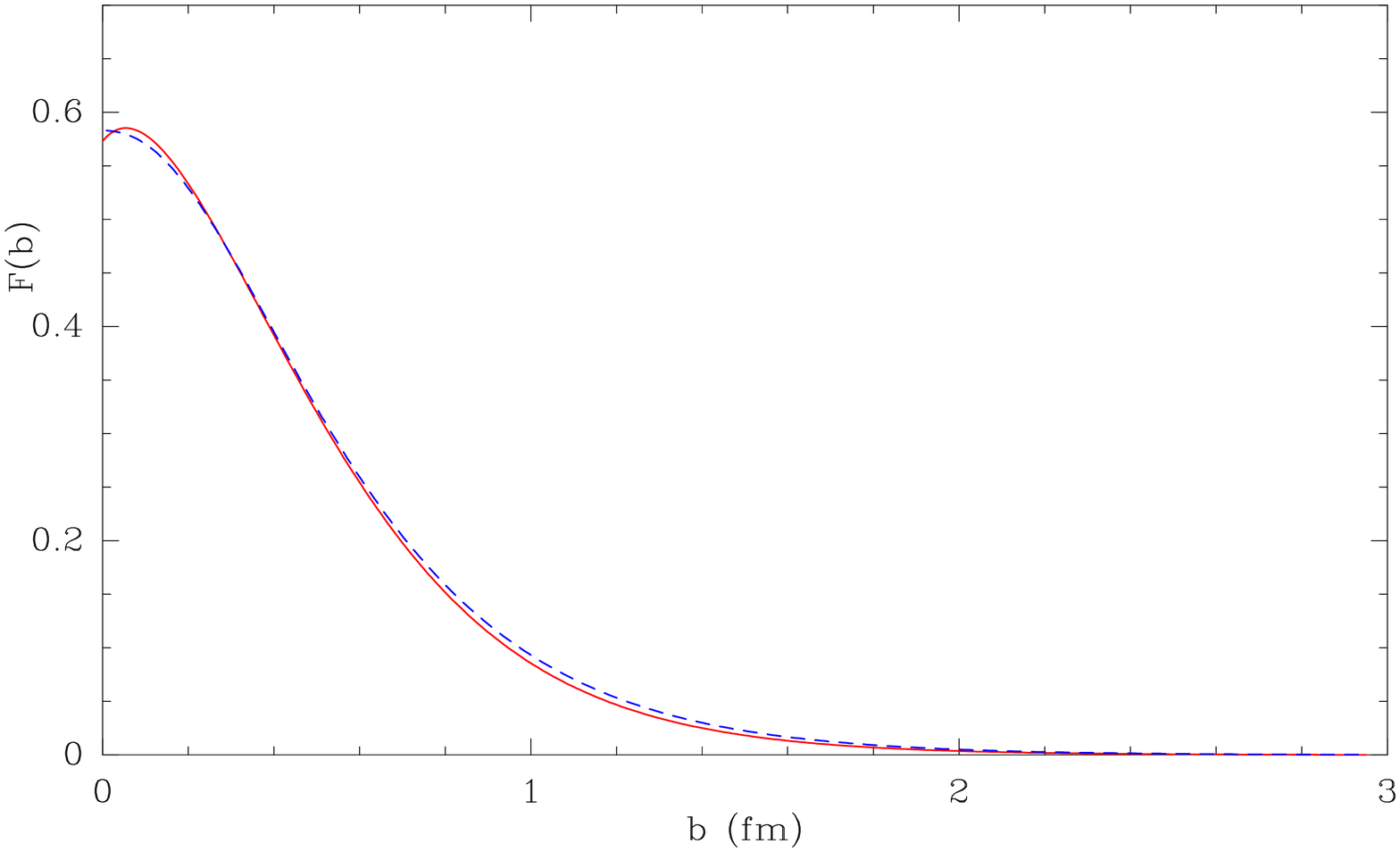,width=12.0cm}
\caption{The profile function $F(b)$ as a function of $b$ for
$\pi^{\pm}~p$ .
 Fermi solid red curve, BSW dashed blue curve.}
\label{fig2pi}
\vspace*{-1.5ex}
\end{center}
\end{figure}

\begin{figure}[htb]
\vspace*{-18mm}
\begin{center}
  \epsfig{figure=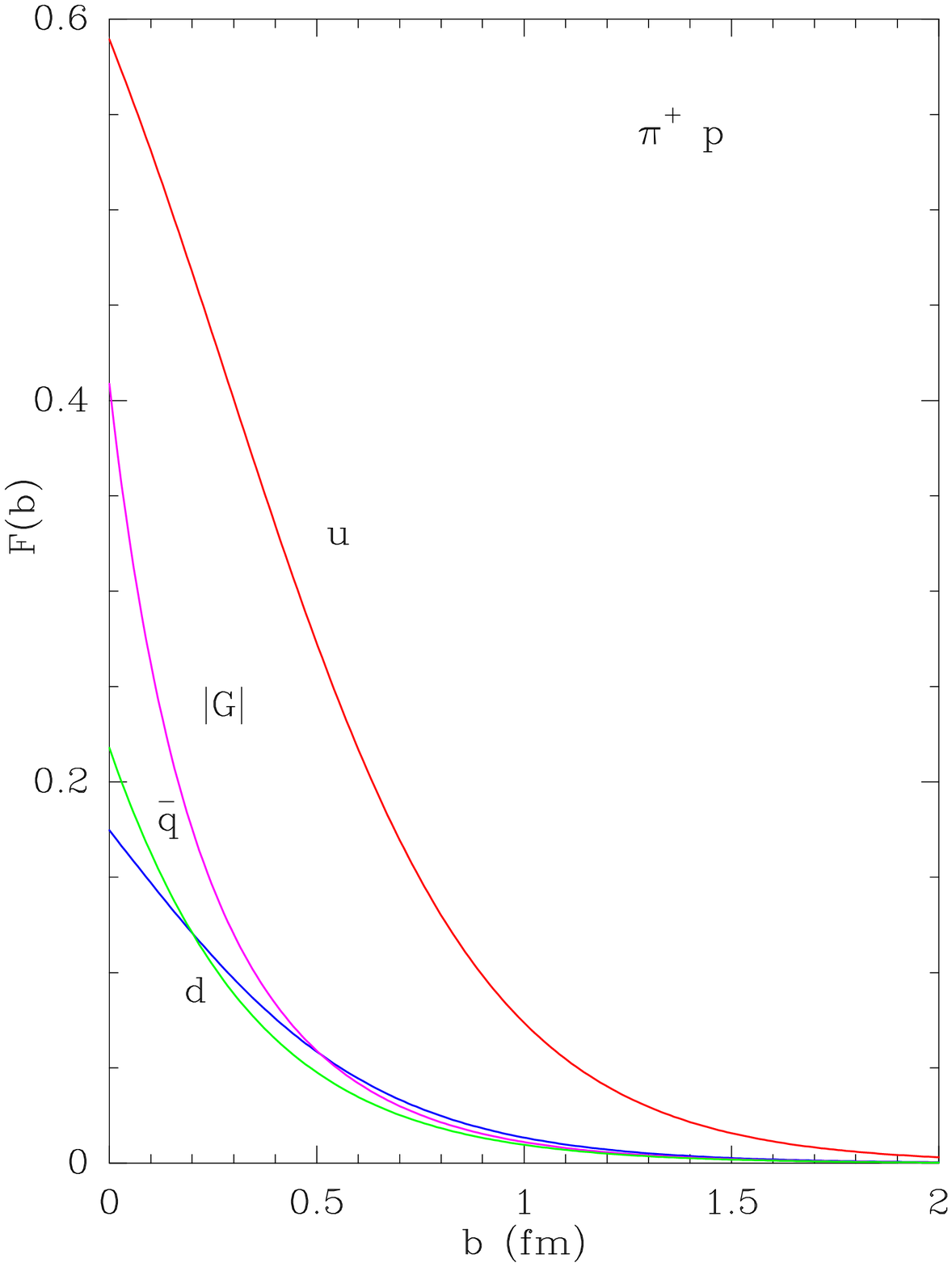,width=7.0cm}
\end{center}
  \vspace*{-5mm}
\caption[*]{Individual contribution of quarks to the profile function
for $\pi^{\pm}~p$ .}
\label{fi0pi}
\vspace*{-0.5ex}
\begin{center}
  \epsfig{figure=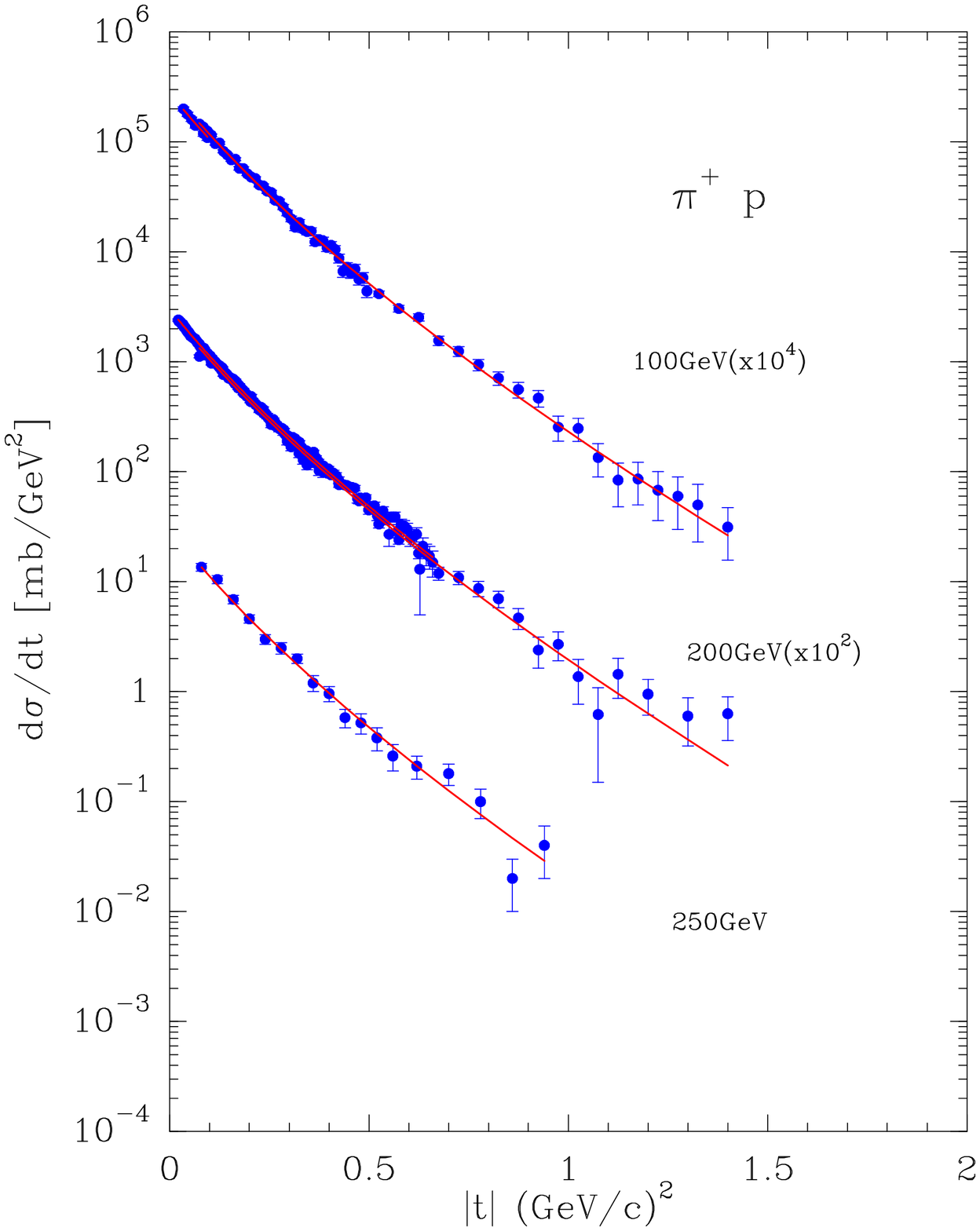,width=8.0cm}
\end{center}
  \vspace*{-5mm}
\caption[*]{The $\pi^{+}~p$ differential cross section as a
function of $|t|$.
Experiments from Refs.  \cite{aker76}-\cite{adam87}.}
\label{fi1pi}
\vspace*{-1.0ex}
\end{figure}

Notice that the parameters $d_0, d_1, d_2, d_3, b_0, Z_{\bar q}, Z_g$ 
are the same for both reactions.
In Figs. \ref{fig1pi}-\ref{fig2pi} a plot is made for $\tilde F(t)$ and
$F(b)$ with a comparison to the BSW profile, the curves are very close
which shows the validity of the Fermi profile.
The variation of $\tilde F(t)$   in $\pi~p$ for BSW shows
a zero at $|t| = 5.6~\mbox{GeV}^2$ while for Fermi the zero 
occurs at $|t| = 6.9~\mbox{GeV}^2$,
this difference in the zero position reflects the dominance of the gluon
over the sea as seen in Fig. \ref{fi0pi}.

Compared to $p~p$ scattering we have not the same range of high energy data so
the pomeron parameters are subject to less constraints, nevertheless it is 
interesting to determine the size of the different components in 
Eq. (\ref{fermpi}). 
With the parameters of Table \ref{tab:table3}
we plot in Fig. \ref{fi0pi} the individual contribution of the
components in the $\pi^+~p$ case,
we observe the dominance of the quark $u$ and the gluon, but the sea 
contribution which was small in $p~p$ (see Fig. \ref{fi0}) becomes more 
sizeable which is expected due to the pion effect.

 We have introduced In Eq. (\ref{fermpi}) the potentials
$X^{\pm}_{u},~X^{\pm}_{d}$ in order to separate the reactions $\pi^{\pm}$
leading to two separated profiles
$F^{\pm}_{\pi}(b)$, with the parameters of
Table \ref{tab:table3} the numerical difference between $F^{+}_{\pi}(b)$ 
and $F^{-}_{\pi}(b)$ is very small, this fact
can be explained by the experimental  the differential cross
sections for the two processes which are close in the  energy range considered 
here, we remark that the difference is in part due to the Regge $\rho$ 
contribution.

\begin{figure}[htb]
\vspace*{-18mm}
\begin{center}
  \epsfig{figure=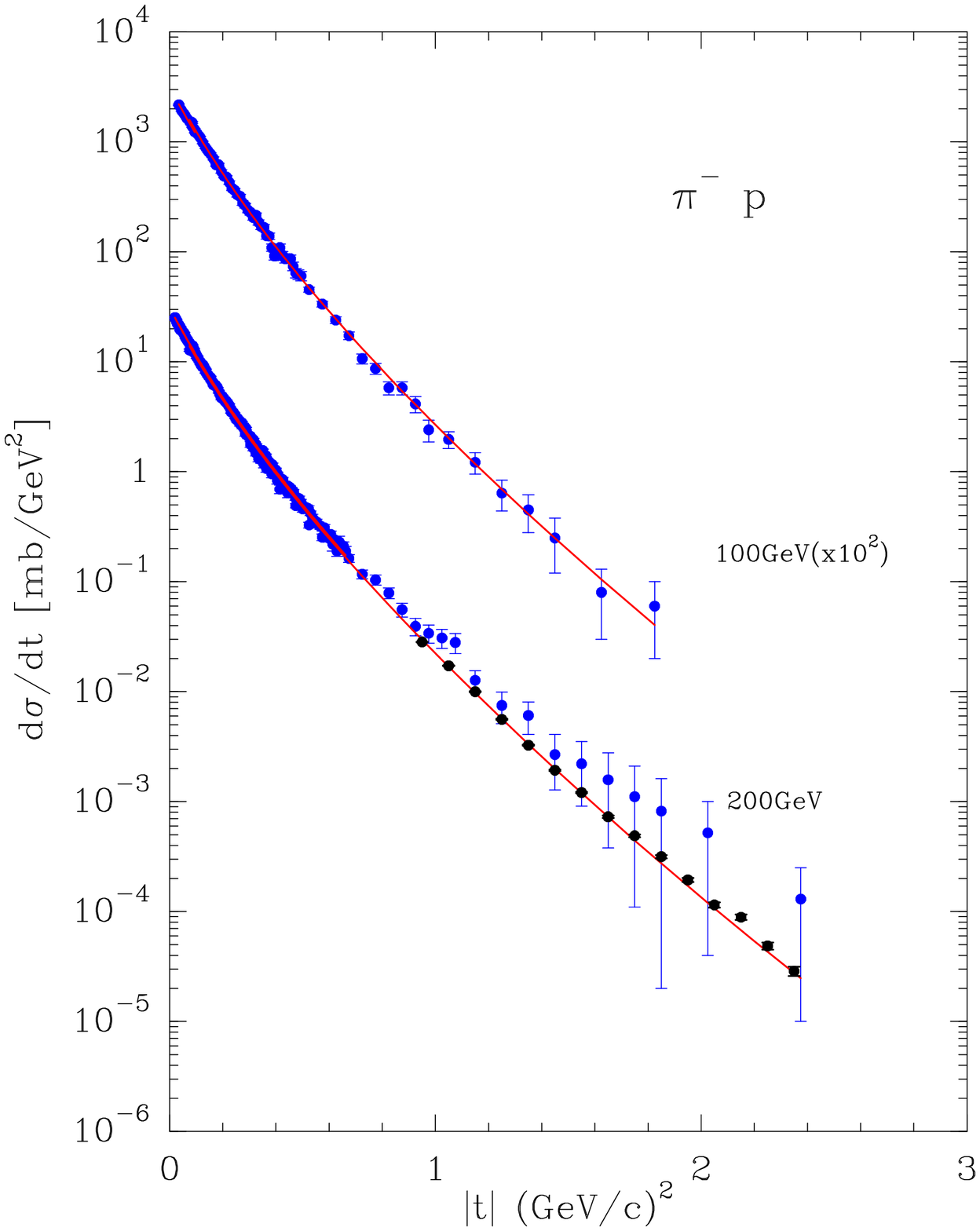,width=8.0cm}
\end{center}
  \vspace*{-5mm}
\caption[*]{The $\pi^{-}~p$ differential cross section as a
function of $|t|$. Experiments from Refs.
\cite{aker76}-\cite{schiz79}.}
\label{fi2pi}
\vspace*{-1.0ex}
\begin{center}
  \epsfig{figure=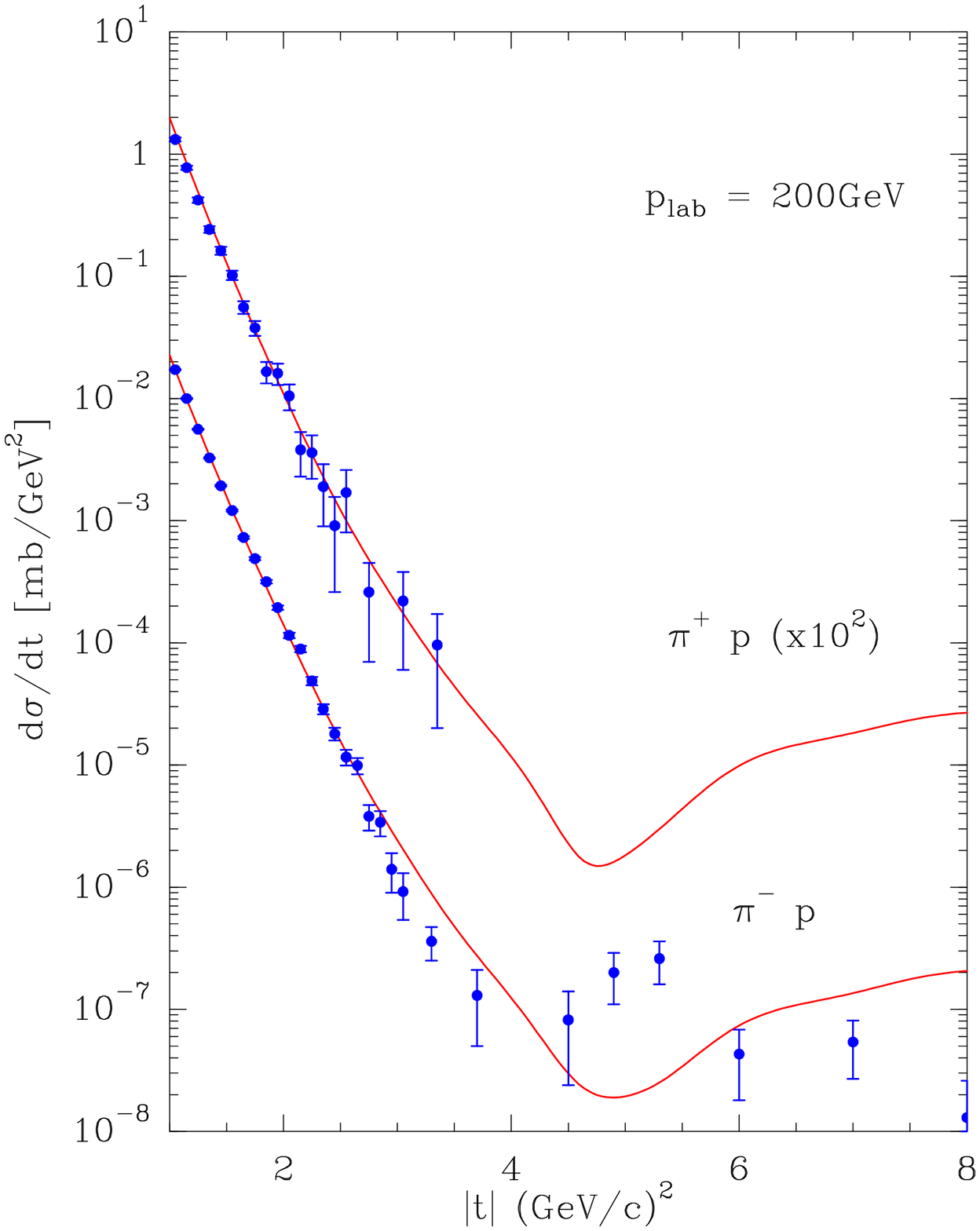,width=8.0cm}
\end{center}
  \vspace*{-5mm}
\caption[*]{The $\pi^{\pm}~p$ differential cross section
for large $|t|$ values. Experiment from Ref. \cite{rubi84}}
\label{fi3pi}
\vspace*{-1.0ex}
\end{figure}

In Figs \ref{fi1pi}-\ref{fi2pi} a plot of the differential cross
sections shows a reasonable agreement with the data. Also, the large
$|t|$ values presented in Fig. \ref{fi3pi} reveal the existence of a dip
around $|t| = 4.5~\mbox{GeV}^2$ consistent with the data.
For the total cross sections we obtain at $p_{lab} = 310$ GeV a value
$\sigma_{tot} = 24.86 \pm 0.2$ mb for $\mathbf{\pi^{-}}~p$ and
$\sigma_{tot} = 24.48 \pm 0.3$ mb for $\mathbf{\pi^{+}}~p$,
the experimental values are respectively 
$\sigma_{tot} = 24.9 \pm 0.08$ mb and $\sigma_{tot} = 24.5 \pm 0.1$ mb
from Ref.  \cite{caroll79}.
Since we have a different pomeron potential for $\pi^-$ and
$\pi^+$ what is the incidence on the total cross section at high energy,
a prediction at $\sqrt{s} = 7$ TeV gives respectively for the two reactions
58.8 mb and 58.2 mb,
 the difference is 1\%, so the near equality of the cross sections at
high energy is preserved in accordance with the Pomeranchuk theorem.


\clearpage
\newpage

\section{The $\mathbf{\pi^{\pm}}~^4\mbox{He}$ elastic scattering}
\label{pihescat}
To study this reaction we follow the same approach  as in the previous sections, 
namely, we keep the parameters c, c' and the thermodynamical potentials to be
identical to those of the $\pi~p$ case.
The parameters are determined from a fit of the CERN data  \cite{burq81} for 
$50 \le E_{lab} \le 300$ GeV and a $|t|$ domain between 0.0086-0.0481 GeV$^2$.
We obtain  a $\chi^2 = 640$ for 584 pts or a $\chi^2/pt = 1.1$
which is close to the pion-proton value.
The obtained parameters for the pomeron are given in Table \ref{tab:table5}:
\begin{table}[hp]
    \centering
    \begin{tabular}{|rllrll|}\hline
$d_0$ & = & $0.072\pm 0.007$, & \;\;$d_1$&=& $17.98699\pm 1.03$\\
$d_2$&=& $23.262\pm 1.20$,&\;\;$d_3$&=& $46.306\pm 2.215$ \\
$b_0$&=& $0.5640\pm 0.0124$ fm&\;\;&&\\ \hline
    \end{tabular}
    \caption{\label{tab:table5} Pomeron parameters of the  Fermi model
for $\pi^{\pm}~^4\mbox{He}$  elastic scattering.}
      \end{table}

\begin{figure}[htb]
\vspace*{-18mm}
\begin{center}
  \epsfig{figure=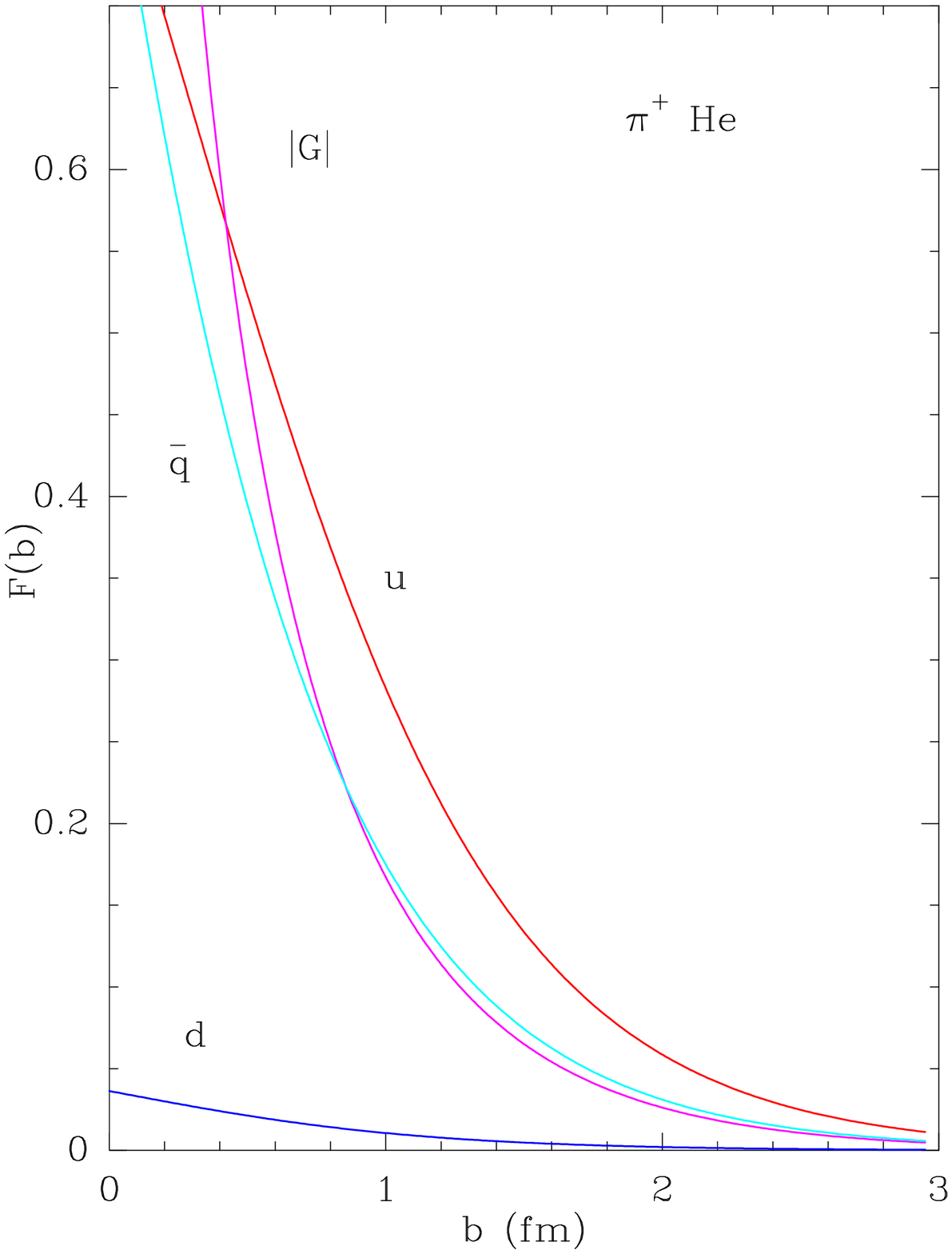,width=7.5cm}
\end{center}
  \vspace*{-5mm}
\caption[*]{Individual contribution of quarks to the profile function
as a function of $b$ for $\mathbf{\pi^{+}}~^4\mbox{He}$  .}
\label{fipiphe0}
\vspace*{-1.0ex}
\begin{center}
  \epsfig{figure=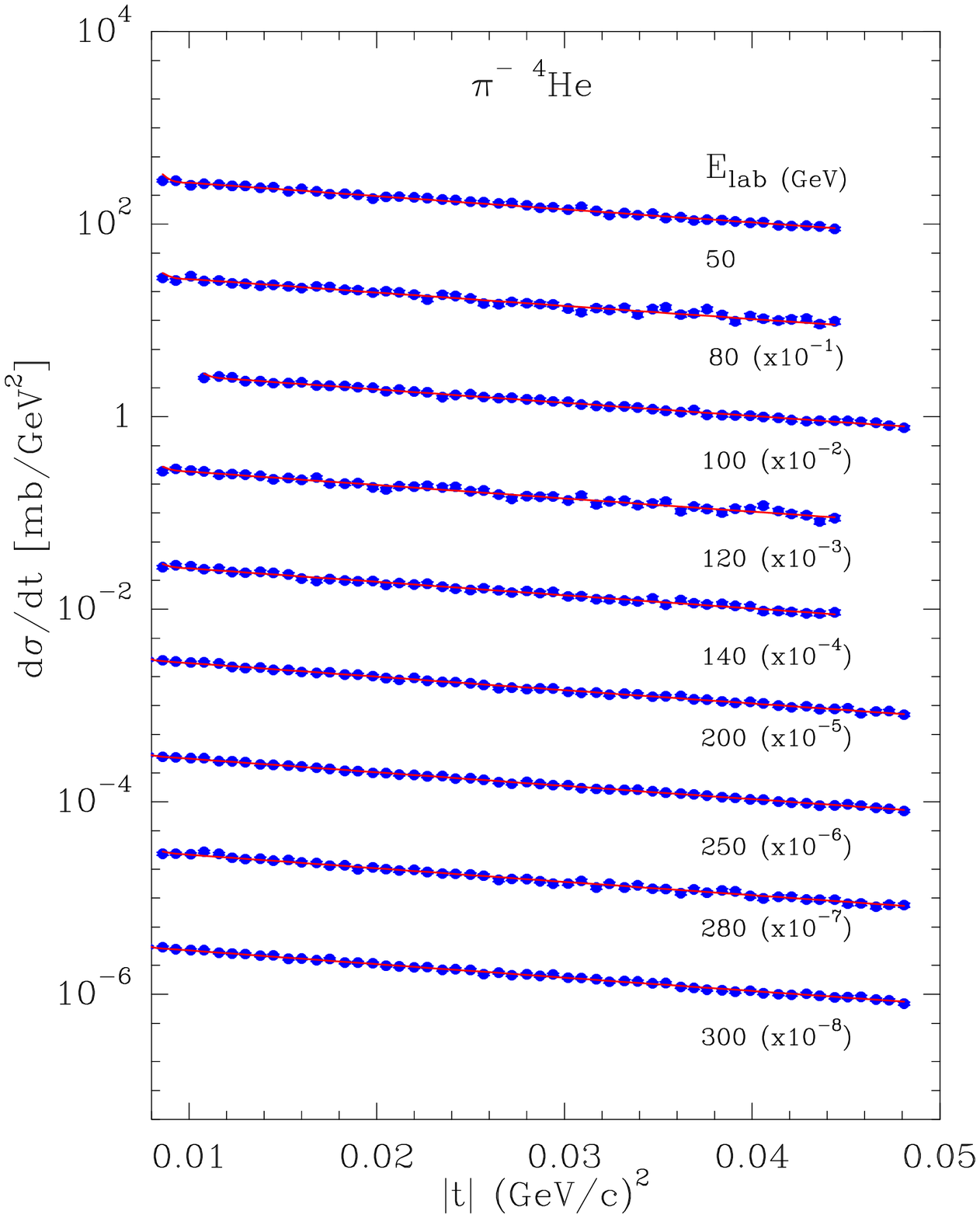,width=8.0cm}
\end{center}
  \vspace*{-5mm}
\caption[*]{The $\mathbf{\pi^{-}}~^4\mbox{He}$ differential cross
section as function of $|t|$. Experiment from Ref.  \cite{burq81}.}
\label{dsigpimhe}
\vspace*{-1.0ex}
\end{figure}

The parameter $b_0$ has the same order of magnitude 
as the one obtained in $p~^4\mbox{He}$.
The different quarks components are plotted in Fig. \ref{fipiphe0}, 
we see that the gluon, the quark u and the sea give the major contributions.

\begin{figure}[htb]
\vspace*{-18mm}
\begin{center}
  \epsfig{figure=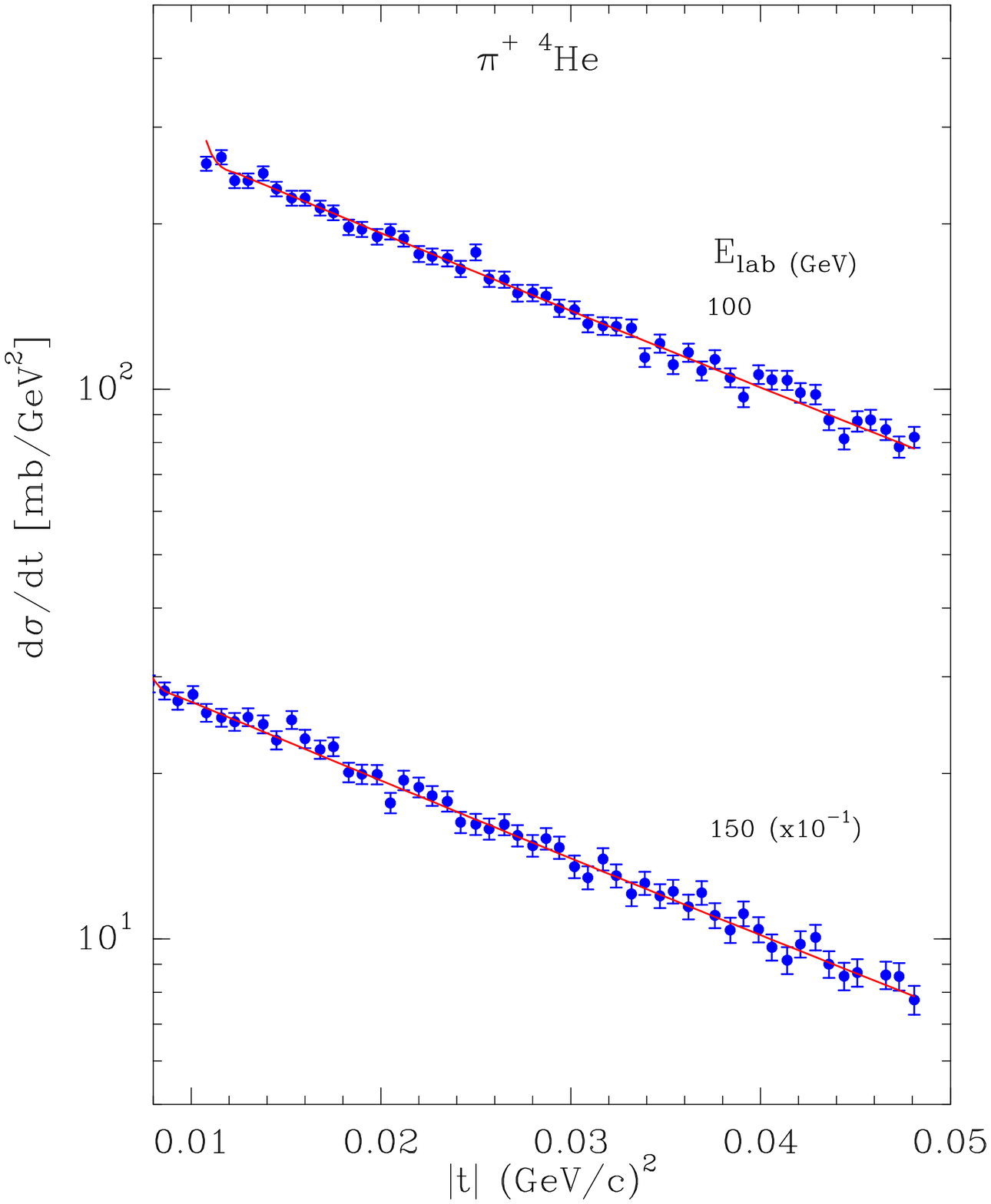,width=8.0cm}
\end{center}
  \vspace*{-5mm}
\caption[*]{The $\mathbf{\pi^{+}}~^4\mbox{He}$ differential cross
section as function of $|t|$.
Experiment from Ref.  \cite{burq81}.}
\label{dsigpiphe}
\vspace*{-1.0ex}
\begin{center}
  \epsfig{figure=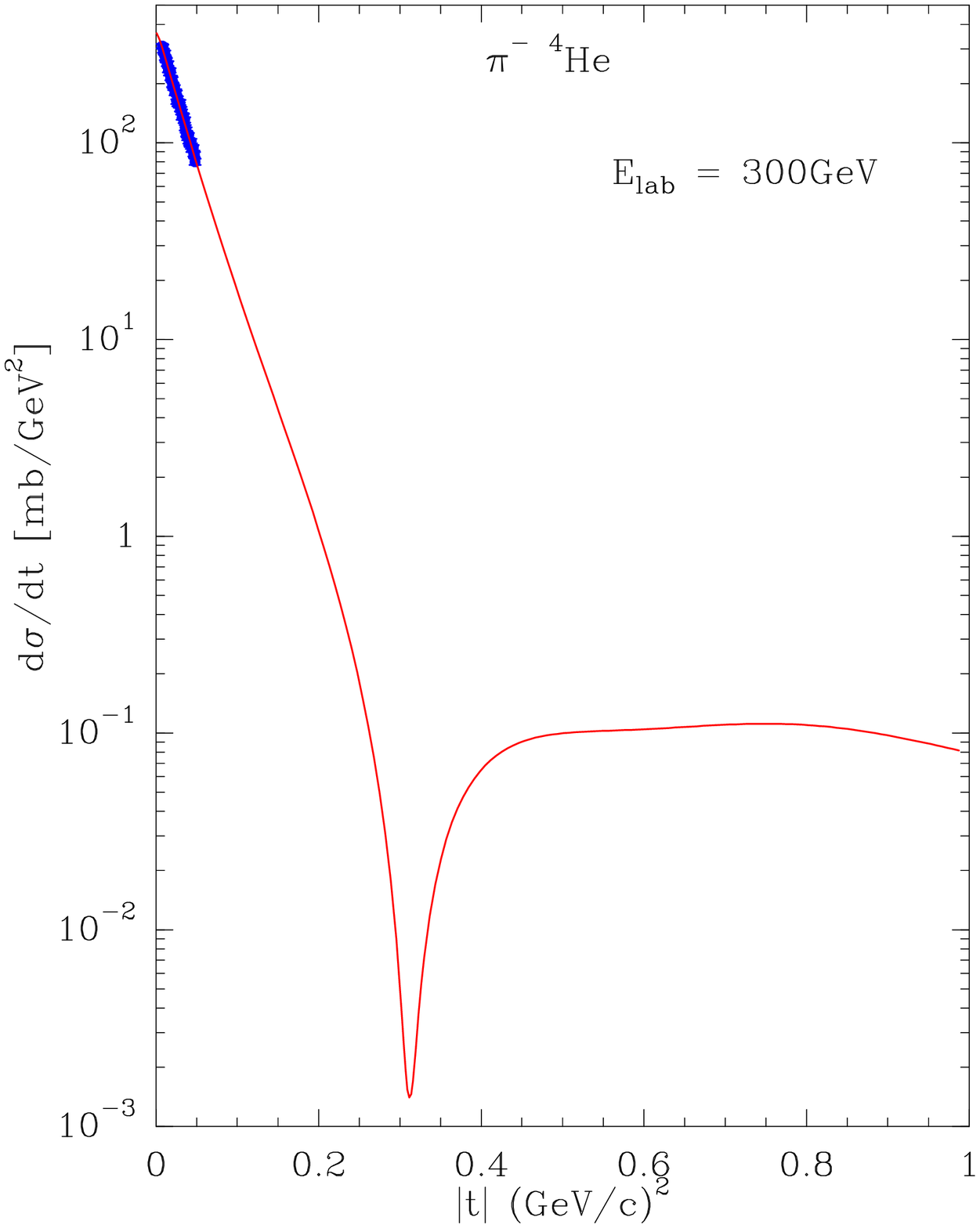,width=8.0cm}
\end{center}
  \vspace*{-5mm}
\caption[*]{The $\mathbf{\pi^{-}}~^4\mbox{He}$ differential cross
section at large $|t|$ values.
Experiment from Ref.  \cite{burq81}.}
\label{dsigpimdip}
\vspace*{-1.0ex}
\end{figure}

We show in Figs. \ref{dsigpimhe}-\ref{dsigpiphe} a plot of
differential cross sections, although the $t$ range is limited to the 
forward direction the agreement with data remains good. 
In Fig. \ref{dsigpimdip} we make a prediction
for the large $|t|$ $\mathbf{\pi^{-}}~^4\mbox{He}$ differential cross section 
at the highest measured  energy  300 GeV,
a dip occurs at $|t| = 0.3~\mbox{GeV}^2$, which is slightly shifted 
to higher $|t|$ value compared to the  reaction 
$\mbox{p}~^4\mbox{He}$ (see Fig. \ref{fiphe1}).

\begin{figure}[hbp]
\begin{center}
  \epsfig{figure=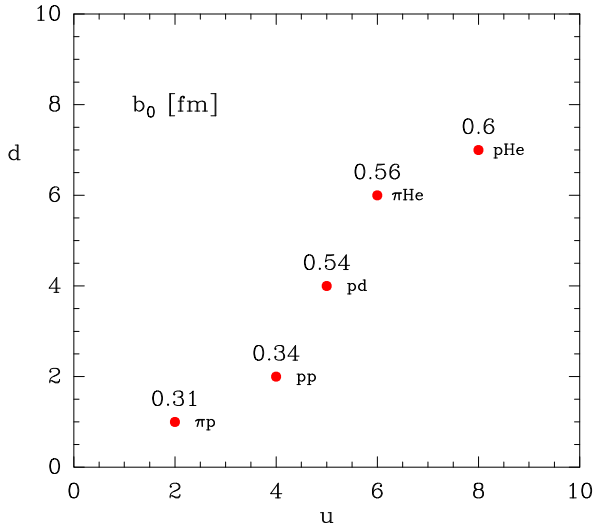,width=9.0cm}
\caption{The $b_0$ values as function of $u$ and $d$ quarks number.}
\label{qdistr}
\end{center}
\end{figure}

For the total cross sections we obtain at $E_{lab} = 150$ GeV a value
$\sigma_{tot} = 83.6 \pm 0.2$ mb for $\mathbf{\pi^{-}}~^4\mbox{He}$ and
$\sigma_{tot} = 85.17 \pm 0.3$ mb for $\mathbf{\pi^{+}}~^4\mbox{He}$,
the experimental values are respectively 
$\sigma_{tot} = 83.0 \pm 0.9$ mb and $\sigma_{tot} = 85.3 \pm 0.7$ mb
from Ref. \cite{burq81}.
\clearpage
\newpage
In the previous sections we made an analysis of 8 reactions,
$p~p, \bar p~p, p~d, p~^4\mbox{He}$, $\pi^{\pm}~p,$  $\pi^{\pm}~^4\mbox{He}$,
the parameter $b_0$ introduced in the profile function
 Eqs. (\ref{ferm}),(\ref{fermpi})
is related to the average size of the interacting partons system. 
 We show in Fig. \ref{qdistr}  a plot of the $b_0$ values as a function 
of the number of quarks $u$ and $d$ which are involved in a reaction, 
we observe an increase of the $b_0$ values with the number of quarks,
an expected feature but interesting to confirm. This result is
similar to the well known nuclear situation where the mean radius of a 
nucleus increases with the corresponding atomic mass number.
\section{Conclusion}
The introduction of Fermi-Dirac functions as a new opaqueness 
built in with different parton components in
impact parameter space gives a reasonable description of 
 8 elastic reactions $p~p$, $\bar p~p$,  $p~d$, $\mbox{p}~^4\mbox{He}$, 
$\pi^{\pm}~p$  and $\pi^{\pm}~^4\mbox{He}$.
The size and the behavior of these components in impact parameter space 
agrees with what we expect in their localization inside the interaction
domain. 
This first simplified approach certainly needs a more refined version  
by the introduction of heavy quarks and also by reducing the number of 
parameters.

We would like to emphasize that
we do not have to rely on the assumption of proportionality between 
the matter distribution
and the charge distribution which was introduced arbitrarily
in the original BSW  because in our Fermi approach the relation 
is obtained in a natural way.
In BSW the presence of the extra term in $\tilde F(t)$ to cancel
a second dip which was never justified is now explained by the role of the
gluon.
We have also proven that the thermodynamical potentials associated to the
partons and determined from the basic interactions in  $p~p$ and $\pi~p$
elastic scattering
are an intrinsic property of the partons also valid for elastic   
light nuclei reactions.

With the same approach one could envisage an extension to the spin amplitudes,
where for each parton one defines two potentials related to the spin 
orientation up-down, in an analogous way to the polarized PDF \cite{BBS1}.
However, due to the scarce measurements of polarized
elastic reactions at high energy there exists  a difficulty 
to obtain reliable values of the parameters.\\

I am grateful to J. Soffer for constructive comments in the preparation
of the manuscript.

\end{document}